\documentclass[12pt]{nature}
\usepackage{amsmath,amsfonts,amssymb,amsthm,epsfig,array}
\usepackage{dsfont,color}
\usepackage{microtype} %%% To avoid \hbox overfull
\usepackage[final]{pdfpages}
\usepackage{braket}
\usepackage[colorlinks,citecolor=blue,linkcolor=blue,urlcolor=blue,hypertexnames=false]{hyperref}
\newcommand{\figref}[2]{\hyperref[#1]{\ref*{#1}#2}}
\usepackage{multirow}
\usepackage[titletoc,title]{appendix}
\makeatletter 
\let\saved@includegraphics\includegraphics
\AtBeginDocument{\let\includegraphics\saved@includegraphics}
\renewenvironment{figure}{\@float{figure}}{\end@float}
\makeatother

\newcommand{\eg}{\textit{e}.\textit{g}.\ }

\title{Enantiomer-Selective Magnetoresistance in Chiral Gold Nanocrystals by Magnetic Control of Surface Potentials}

\author{ \center Fengxia Wu$^{1~\ast}$, Ying Wang$^{2~\ast}$, Yufei Zhao$^{3~\ast}$, Zhenyu Yang$^{6}$~\footnote{These authors contribute equally.}, Yu Tian$^1$, Zuoti Xie$^{5,6~\dag}$, Wenxin Niu$^{1~\dag}$, Binghai Yan$^{3,4~\dag}$, Cunlan Guo$^{2}$ \footnote{Correspondence: zuoti.xie@gtiit.edu.cn (Z.X.), niuwx@ciac.ac.cn (W.N.), binghai.yan@weizmann.ac.il (B.Y.), cunlanguo@whu.edu.cn (C.G.)}  \\
\normalsize{ 
$^1$ State Key Laboratory of Electroanalytical Chemistry, Changchun Institute of Applied Chemistry, Chinese Academy of Sciences, Changchun 130022, China; School of Applied Chemistry and Engineering, University of Science and Technology of China, Hefei 230026, China \\
$^2$ College of Chemistry and Molecular Sciences, Wuhan University, 299 Bayi Road, Wuhan, Hubei 430072, China \\
$^3$ Department of Condensed Matter Physics, Weizmann Institute of Science, Rehovot 76100, Israel \\
$^4$ Department of Physics, the Pennsylvania State University, University Park, PA, 16802, USA \\
$^5$ Department of Materials Science and Engineering, MATEC, Guangdong Technion-Israel Institute of Technology, Shantou, Guangdong 515063, P. R. China; \\
$^6$ Quantum Science Center of Guangdong-Hong Kong-Macao Greater Bay Area (Guangdong), Shenzhen-Hong Kong International Science and Technology Park, Shenzhen, Guangdong 518000, P. R. China }
}

\usepackage{mathptmx} 
\begin{document} 
\maketitle 

\begin{abstract}
Chiral nanomaterials offer intriguing possibilities for novel electronic and chemical applications. Here, we report the discovery of an enantiomer-selective magnetoresistance effect in chiral gold nanocrystals. Based on precise control of nanocrystal chiral morphology using amino acid-directed synthesis, we demonstrate that an external magnetic field can dramatically modulate resistance in an enantiomer-specific manner. For a given enantiomer, a magnetic field in one direction alters the resistance by dozens of times, while the opposite field direction leaves it unchanged. 
This asymmetric response reverses for the opposite enantiomer \textcolor{black}{
and are reproduced in both single nanocrystals by conduction atomic force microscopy and nanocrystal thin films in solid state devices. } We attribute this phenomenon to a chirality-driven charge pumping effect, where the interplay between the chiral morphology and the magnetic field selectively modifies the surface potential. The magnitude and sign of the magnetoresistance can be further tuned by the surface chemistry of the nanocrystal, as demonstrated through sulfide treatment. Our findings reveal a new form of chirality-dependent magnetoresistance, distinct from previously known effects such as chirality-induced spin selectivity and electric magnetochiral anisotropy.
The ability to remotely control surface potentials of chiral nanostructures using magnetic fields could enable novel approaches in catalysis, drug delivery, and nanoelectronics.
\end{abstract}

\section{Introduction}
The exploration of chirality has unlocked exciting possibilities for developing unconventional nanomaterials with distinctive properties and diverse applications, including chiroplasmonics \cite{Hentschel2017}, chiral catalysis \cite{kobayashi1999catalytic}, 
spintronics \cite{yang2021chiral}, enantiomer discrimination \cite{maier2001separation}, and enantiomer-dependent chemical and biological responses \cite{xu2022enantiomer}. 
Chiral quantum materials are gaining rapidly increasing research interest due to their fascinating magneto-transport properties, such as the chirality-induced spin selectivity (CISS)~\cite{Naaman2012,Naaman2019} and electric magnetochiral anisotropy (EMCA)~\cite{Rikken2001}.

In CISS, chiral organic molecules (e.g., DNA or peptides) are sandwiched between a ferromagnetic electrode and a metal, exhibiting dramatically different resistance upon flipping the electrode's magnetization~\cite{Xie2011,Liu2022experiment,Al-Bustami2022}. Although the mechanism of CISS magnetoresistance (MR) is still under debate~\cite{evers2021theory,Zhao2025,Liu2023spin,yan2024structural}, the ferromagnetic electrode, rather than a magnetic field, is believed to be essential for generating observable MR. In EMCA, chiral inorganic or organic crystalline conductors~\cite{pop2014electrical,Aoki2019,Rikken2019,Inui2020,Shiota2021,Ye2022Te} exhibit current direction-dependent MR in an external magnetic field. Furthermore, EMCA adheres to Onsager's reciprocal relation by giving the same zero-bias resistance at opposite magnetic fields, while CISS MR violates reciprocity by showing significantly different zero-bias resistance for opposite magnetizations in the electrode \cite{rikken2023comparing,yan2024structural}.

The chirality commonly originates in the molecule or crystal structure in these studies. Recently, inorganic materials with chiral morphology, for example, chiral gold nanocrystals (Au NCs)~\cite{lee2018amino,gonzalez2020micelle}, have been synthesized in a large scale and garnered considerable attention because they bring novel modes for molecular recognition and biological regulation~\cite{Tang2022review,Kotov2023review}. For example, chiral Au NCs with strong chiroplasmonic optical activity can generate superchiral electromagnetic fields for enantioselective sensing of adsorbed amino acids and proteins~\cite{kim2022enantioselective}. The high dissymmetry factor of chiral Au NCs enables the design of efficient circularly polarized light-detecting transistors \cite{namgung2022circularly} and sensitive chiroplasmonic hydrogen sensors~\cite{lv2023engineering}. Furthermore, the chiral selectivity between chiral molecules and chiral Au NCs also results in distinguishable electrochemical behaviors for enantioselective catalysis and chiral discrimination \cite{wu2022synthesis,wu2023surface}. Notably, chiral Au NCs can regulate the maturation of immune cells and exhibit distinct immune responses by their chirality \cite{xu2022enantiomer,KunLiu2022CCS}.
Known the versatile surface chemistry, the magneto-transport of chiral Au NCs will provide deep insights into controlling these chirality-dependent processes by a magnetic field.

In this work, we discovered a unique chirality-driven MR in chiral Au NCs by applying an external magnetic field. 
We synthesized Au NCs with pinwheel-like chiral morphology by an amino-acid-directed strategy and measured the resistance of single NCs using conducting atomic force microscopy (cAFM)
\textcolor{black}{and NC films by solid-state devices.} Distinct from the known CISS or EMCA, the resistance of chiral Au NCs is largely enhanced (positive MR) in one field orientation and nearly unchanged in the opposite field, showing a salient asymmetric feature with respect to the zero-field case. The asymmetric order was reversed when we switched to NCs with opposite handedness, verifying the role of chirality \textcolor{black}{and remains robust against temperature (10-400 K).} By modifying the NC surface capping layer with sulfide, we can also reduce the resistance (negative MR) by one field direction and keep it unchanged in the opposite field. We rationalize the usual MR by the magnetochiral charge pumping~\cite{Zhao2025} in the surface capping layer, which controls the surface potential and the total resistance. 
Our work provides valuable insights to understand chirality-driven transport effects and offers implications for selectively controlling the surface potential and manipulate the surface activity for chemical and biological reactions.

\section{Synthesis and characterization of chiral Au NCs}

\begin{figure}
    \centering
    \includegraphics[width=1\linewidth]{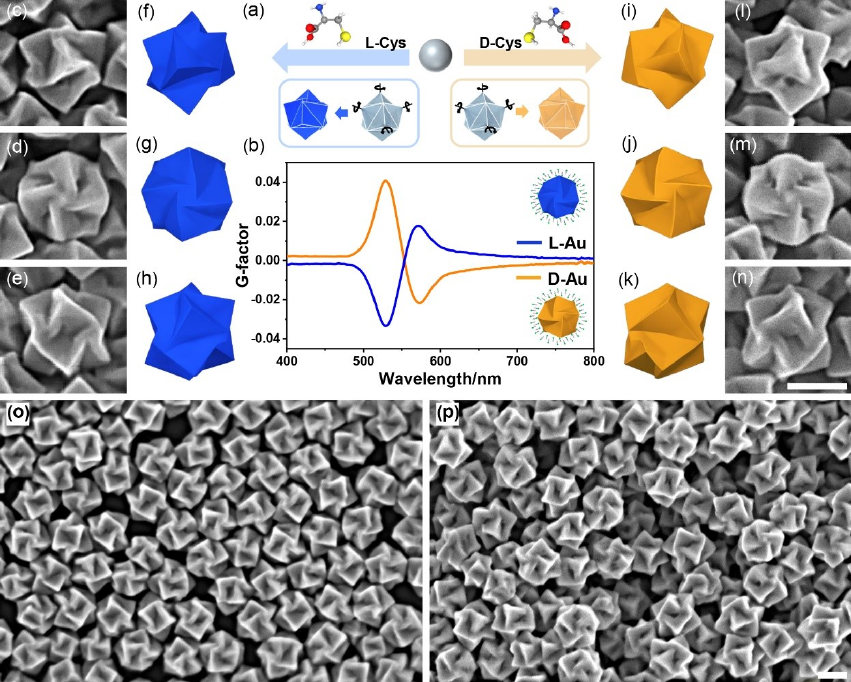}
    \caption{\textbf{Synthesis, characterization, and chiroptical properties of chiral Au NCs.}   (a) Schematic illustration of the processes of Cys-directed growth of chiral Au NCs with opposite chirality. The pyramidal subunits on the trisoctahedron intermediates are rotated clockwise or counterclockwise. (b) The g-factor spectra. The insets show corresponding geometrical models of the chiral Au NCs with CTAB bilayers. (c-n) SEM images and corresponding geometric models of individual chiral Au NCs synthesized with L-Cys (c, d, e, f, g, h) and D-Cys (i, j, k, l, m, n) as chiral inducers, respectively. SEM images and geometric models in (c, f, i, l), (d, g, j, m), and (e, h, k, n) are viewed from the $\braket{111}$, $\braket{100}$, and $\braket{110}$ axes of the corresponding NCs, respectively. Note that the CTAB bilayers were omitted from the models in (f-k) for clarity. (o, p) Low-magnification SEM images of chiral Au NCs synthesized from L-Cys (o) and D-Cys (p), respectively. The L-Au and D-Au NCs with mean sizes of $150.2 \pm 4.8$ nm and $154.4 \pm 3.6$ nm were synthesized with 80 $\mu L$ of seeds. Scale bars: 100 nm in (n) and (p). (c-e, l-n) share the same scale bar in (n). (o) shares the same scale bar in (p).
    }
    \label{figure1}
\end{figure}

The Au NCs with intrinsic chirality were synthesized based on an amino-acid-directed chirality transfer strategy~\cite{lee2018amino,wu2022synthesis} (see Methods and Figure~\ref{figure1}). With L- or D-cysteine (Cys) as chiral inducers, spherical Au seeds were grown into chiral Au NCs with opposite chirality. After the syntheses, the Au NCs were collected by centrifugation and redispersed in water with cetyltrimethylammonium bromide (CTAB) as the stabilizing agent. Note that CTAB (achiral) is adsorbed on Au surfaces to form a core-shell structure~\cite{mosquera2023surfactant}. \textcolor{black}{To clarify, the Au NCs obtained with L- and D-Cys are designated as L-Au and D-Au NCs, respectively, and the achiral Au NC with racemic Cys is designated as A-Au NCs (Extended Data Table~\ref{exTable1})}
The chiral Au NCs synthesized with L- and D-Cys exhibit inverted circular dichroism (CD) spectra with respect to each other, with dissymmetry factors (g-factors) of approximately -0.033 and 0.041, respectively (Figure~\figref{figure1}{b} and Extended Data Figure~\ref{exFigure1}). From the scanning electron microscopy (SEM) images, the pyramid-like subunits on the chiral Au NCs are rotated clockwise or counterclockwise with respect to the achiral trisoctahedral Au NCs obtained with racemic Cys (Figures~\figref{figure1}{c},\figref{figure1}{f},\figref{figure1}{i},\figref{figure1}{l}, and Extended Data Figure~\ref{exFigure2}). The rotation direction of the pyramidal subunits is dictated by the chirality of the chiral inducers, due to the enantioselective interactions of L- or D-Cys with R- or S- chiral Au surfaces~\cite{lee2018amino, wu2022synthesis}. After the rotation, the four adjacent pyramids on the trisoctahedron no longer share one vertex, but instead form a pinwheel-like structure. As shown in Figures~\figref{figure1}{d} and \figref{figure1}{m}, the pinwheel-like structure of the chiral Au NCs is rotated clockwise and counterclockwise, respectively. Furthermore, by adjusting the amounts of seeds in the growth processes, chiral Au NCs can be synthesized with sizes ranging from 150.2 to 475.5 nm with tunable chiroplasmonic properties (Extended Data Figures~\ref{exFigure3} and ~\ref{exFigure4}, and Extended Data Table~\ref{exTable2}). \textcolor{black}{
We further characterized the magnetic properties of chiral and achiral Au NCs by superconducting quantum interference device (SQUID) measurements. Independent from chirality, these Au NCs exhibit extremely weak ferromagnetism {(Extended Data Figure~\ref{exFigure25})}, similar to earlier observation in Au nanoparticles\cite{nealon2012magnetism}.
} 

\section{Magnetoresistance of chiral Au NCs}

\begin{figure}
    \centering
    \includegraphics[width=0.9\linewidth]{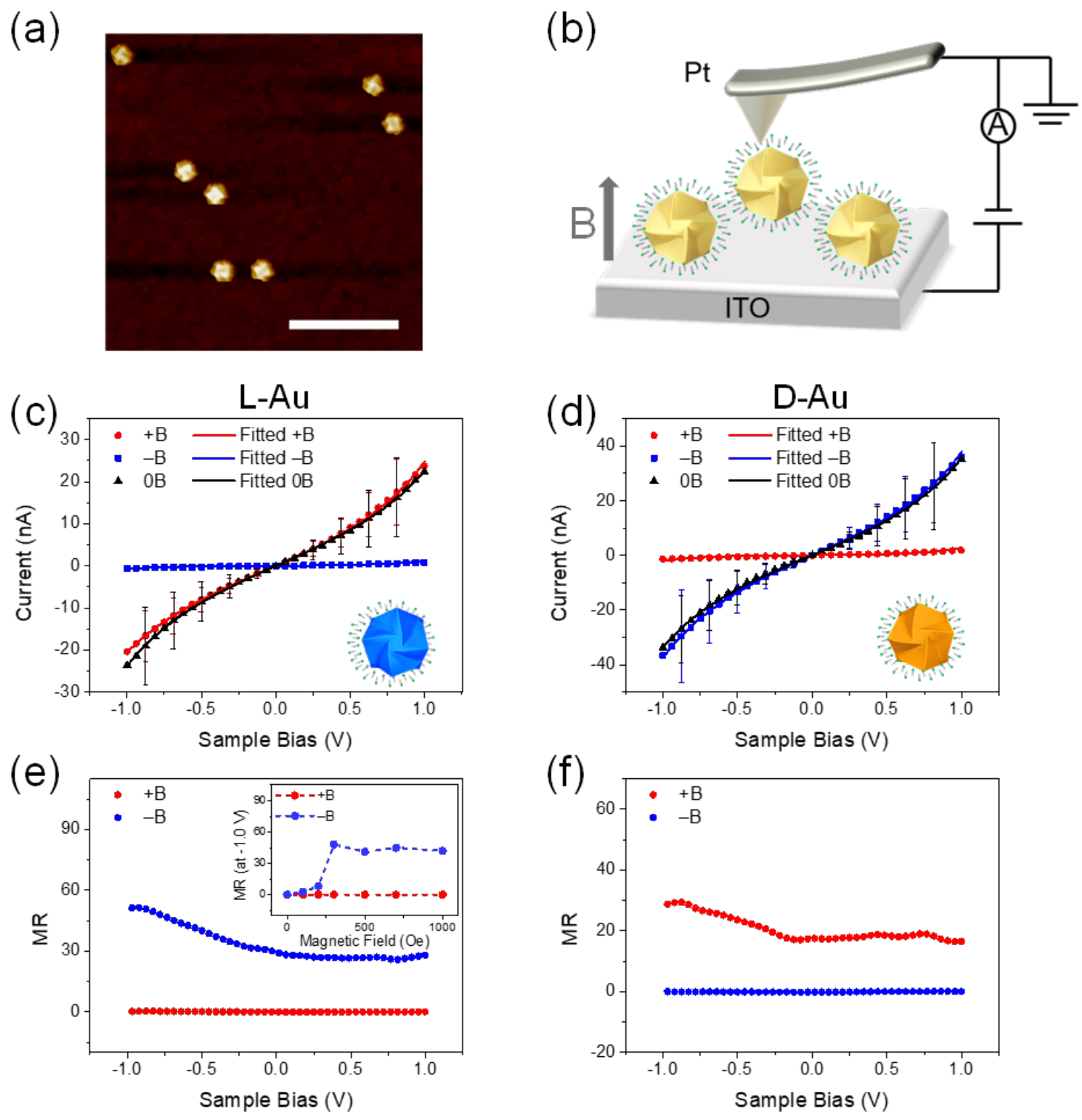}
    \caption{\textbf{Magneto-transport and asymmetric MR for chiral Au NCs.}  (a) An AFM image of D-Au NCs on the indium tin oxide (ITO) substrate. The scale bar is 1 $\mu$m. (b) Schematic of the ITO/chiral Au NC/Pt junction constructed via cAFM. A permanent magnet (300 Oe) is placed beneath the ITO substrate. (c) and (d) The experimental (dots) current-voltage (I-V) curves for the L-Au NC and D-Au NC under different directions of magnetic fields(${\pm\textbf{B}}$). \textbf{0B} (black) stands for zero magnetic field. Solid lines represent fitting using Eq.~\eqref{equation1}. (e) and (f) The MR of the L-Au NC and D-Au NC extracted from (c) and (d), respectively. \textcolor{black}{The inset in (e) shows the changes of MR with magnetic field intensity.} }
    \label{figure2}
\end{figure}

We measured the MR of the chiral Au NCs by cAFM with a Pt tip at room temperature. Using chiral Au NCs with size of $\sim$150 nm as exemplars, we dispersed L- and D-Au NCs on an indium tin oxide (ITO) substrate, enabling MR measurements at the single-nanoparticle level (Figure~\figref{figure2}{}). The size of the L- and D-Au \textcolor{black}{NCs are} $150.2\pm4.8$ and $154.4\pm3.6$ nm, respectively. 
We recorded current-voltage (I-V) relationships of individual chiral Au NCs via the ITO/chiral Au NC/Pt junction. An external magnetic field (\textbf{B} = 300 Oe) was applied orthogonally to the ITO substrate using a permanent magnet placed below the ITO (Figure~\figref{figure2}{b}). Unlike CISS MR studies~\cite{Bloom2024}, our setup did not incorporate a magnetic substrate or tip.
The insulating CTAB molecular shell on Au NCs contributes significantly more to the resistance than the metallic Au NC core~\cite{jin2006gold}, resulting in small currents of tens of nA at 1.0 V bias. Figures \figref{figure2}{c} and \figref{figure2}{d} present the averaged I-V curves for L- and D-Au NCs, respectively, derived from hundreds of individual measurements (see Extended Data Figure~\ref{exFigure5}).
The I-V curves were measured 320--679 times on more than 30 individual Au NCs from at least three parallel samples. The error bar was calculated as the standard deviation of all results. 
We observed an asymmetric MR behavior between \textbf{+B} and \textbf{--B} (Figure~\ref{figure2}). For \textcolor{black}{L-Au NC}, the I-V curve at \textbf{+B} closely resembles the zero-field case within error bars, while at \textbf{--B}, the resistance increases dramatically by 20--50 times. Conversely, for D-Au NCs, \textbf{–B} barely affects the resistance, while \textbf{+B} increases it by $\sim$20 times. \textcolor{black}{The I-V curves of L-Au NC display consistent behavior with those of ITO/L-Au NC/Pt when the Pt AFM tip is replaced with Au tip, indicating that the electrode material does not affect the MR behavior (Extended Data Figure~\figref{exFigure21}). The I-V curves obtained through direct contact of ITO/Pt electrodes are independent of the magnetic field (Extended Data Figure~\figref{exFigure20}).} This reversal of asymmetric behavior between L- and D-Au NCs strongly indicates its chirality-dependent origin.
We quantify MR as the relative change in resistance due to the \textbf{B}-field: MR=$[R(\textbf{B})-R(0)]/R(0)$. The results show that MR is negligible at \textbf{+B} (\textbf{–B}) but substantial at \textbf{–B} (\textbf{+B}) for L(D)-Au NCs, respectively \textcolor{black}{(Figures \figref{figure2}{e} and \figref{figure2}{f}).}

\begin{figure}
    \centering
    \includegraphics[width=1\linewidth]{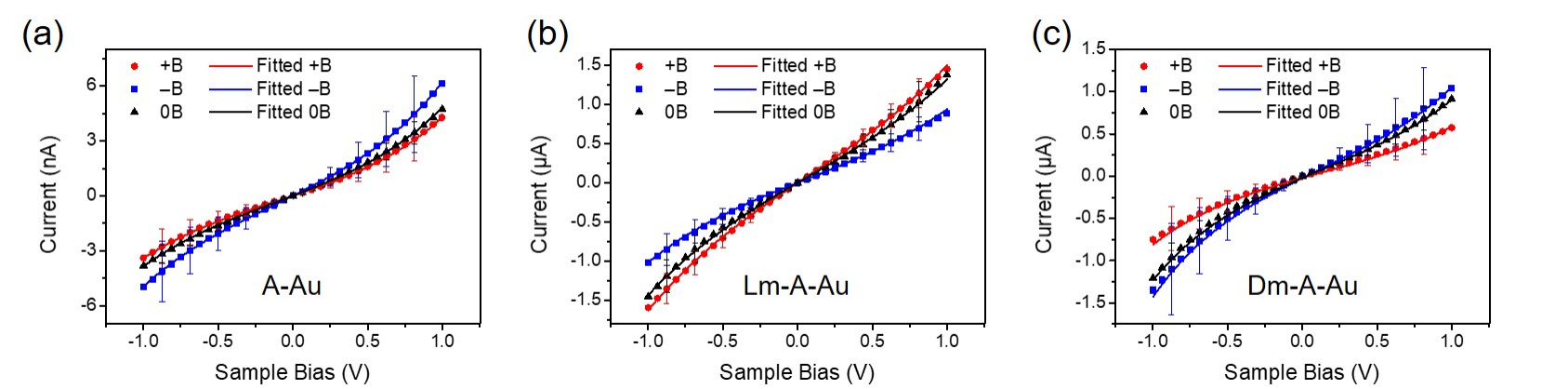}
    \caption{\textbf{Magnetotransport of A-Au, Lm-A-Au, and Dm-A-Au NCs.}   (a)-(c) Current-voltage curves of the A-Au, Lm-A-Au, and Dm-A-Au NCs (from left to right) under magnetic field directions of \textbf{+B} (red), \textbf{–B} (blue), and \textbf{0B} (black), respectively. At least 430 current-voltage curves were obtained for all samples, with approximately 10 measurements performed on each Au NC. The error bar was defined as the standard deviation of all current values measured at each magnetic field condition. \textcolor{black}{The I-V characterizations are performed by cAFM.}}
    \label{figure3}
\end{figure}

To determine whether the chiral morphology of the Au core or the chiral ligand shell is primarily responsible for the observed MR results, we conducted control experiments using three additional types of Au NCs. In addition to the L-Au and D-Au NCs, we investigated achiral Au NCs synthesized with racemic Cys (A-Au, Extended Data Figure~\ref{exFigure2}), L-cysteine-modified achiral Au NCs (Lm-A-Au), and D-cysteine-modified achiral Au NCs (Dm-A-Au). We prepared Lm-A-Au and Dm-A-Au by modifying achiral non-Cys Au NCs with L- and D-Cys, respectively (Extended Data Figures \ref{exFigure6} and \ref{exFigure7}).
Our initial measurements of the I-V behavior of A-Au NCs revealed negligible MR for \textbf{±B} (Figures \figref{figure3}{a} and Extended Data Figure~\ref{exFigure8}). For Lm-A-Au and Dm-A-Au NCs, while the asymmetric order between \textbf{±B} showed a trend similar to L- and D-Au NCs, the magnitude of their MR was significantly smaller (Figure \ref{figure3}). The I-V curves under \textbf{±B} and \textbf{0B} were only marginally or hardly distinguishable when considering error bars for A-, Lm- and Dm-Au NCs. These findings strongly suggest that the substantial MR observed in chiral Au NCs is primarily attributable to the chiral morphology of the NC core rather than the molecular chirality of the ligand shell. This conclusion highlights the crucial role of the NC's intrinsic morphology chirality in determining its MR properties.

Having established the chiral origin of the observed MR, we proceeded to investigate the robustness of this chirality-driven effect under various conditions, including different NC sizes, magnetic field strengths \textcolor{black}{and direction, as well as} AFM forces, to rule out potential artifacts.
Using L-Au NCs as a model system, we observed that the current through chiral Au NCs increases gradually as the NC size increases from 150.2 to 350.5 nm, consistent with previously reported transport behaviors of metal nanoparticles~\cite{jin2006gold} (Extended Data Figures \figref{exFigure9}{a} and \ref{exFigure10}). Concurrently, the overall MR value decreases with increasing NC size, while the MR values under \textbf{–B} consistently remain significantly larger than those under \textbf{+B} (Extended Data Figure~\figref{exFigure9}{d}), confirming the persistence of asymmetric MR across different NC sizes.
\textcolor{black}{ In the magnitude of the asymmetric MR also varies with the magnetic field strength. For L-Au NCs, as the external magnetic field increases from 0 to 1000 Oe, the current at +1.0 V remains nearly constant for \textbf{+B}, while it first decreases and later saturates beyond 300 Oe for \textbf{–B}. Correspondingly, the MR under \textbf{+B}  keeps constant while the MR under \textbf{–B} substantially increases from 0\% to 4800\% from 0 to 300 Oe (Inset in Figure~\ref{figure2}{e} and Extended Data Figures \ref{exFigure9}{b},  \ref{exFigure9}{e}, and \ref{exFigure11}).}
\textcolor{black}{It is noteworthy that when \textbf{B} is perpendicular to the current direction, resistance of chiral Au NCs remains nearly identical to that of \textbf{0B}} {(Extended data Figure \ref{exFigure22})}.
In cAFM measurements, the contact between the AFM tip and Au NCs sensitively influences the total resistance~\cite{Sachi2014bRAFM,Kun2014AFMforce}. To exclude potential artifacts arising from this contact, we evaluated the MR of L-Au NCs under various AFM forces. As anticipated, the current values increase with incremental force due to the compression of the molecular layer and the increased contact area between the tip and the Au NC. Notably, the asymmetric MR remains robust across different applied forces, although the MR at \textbf{–B} exhibits a weakly non-monotonic change with increasing force (Extended Data Figures \figref{exFigure9}{c} and \figref{exFigure9}{f}). This robustness against varying AFM force can be attributed to the stable morphology of crystalline nanoparticles compared to soft organic materials.
\textcolor{black}{
In addition, we exclude that MR is caused by the displacement of Au NCs when applying the magnetic field. Because the magnetic susceptibility shows no chirality dependence (Extended Data Figure~\figref{exFigure25}), the field-induced motion cannot generate chirality-dependent resistance even if it exists. As shown in the following Sec.~\ref{sec:device}, we further fabricated solid-state devices that fix all NCs and reproduced results measured by cAFM.}
These comprehensive investigations across different experimental parameters underscore the intrinsic nature and reliability of the observed chirality-driven magnetoresistance effect in Au NCs. 

\section{Control the sign of magnetoresistance}

\begin{figure}
    \centering
    \includegraphics[width=1\linewidth]{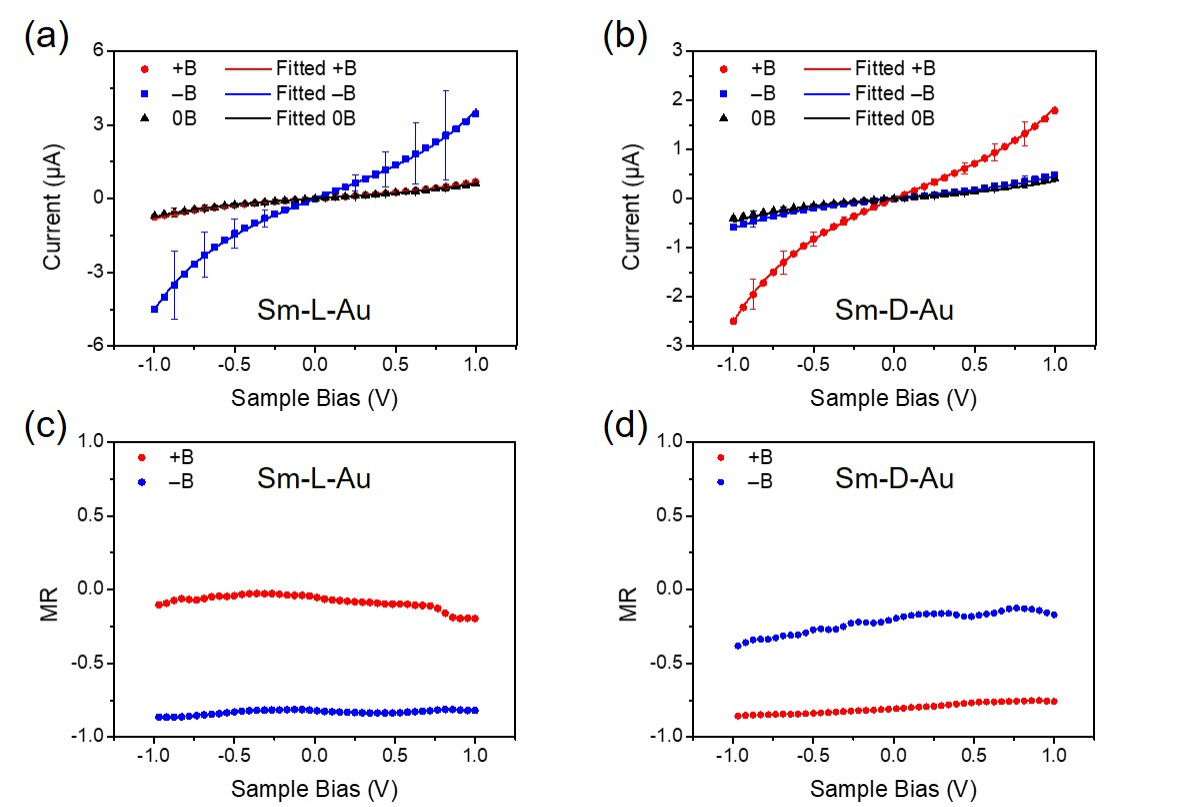}
    \caption{\textbf{
Magnetoresistance characterization of the sulfide-modified chiral Au NCs. }(a) and (b) The experimental (dots) and fitted (line) current-voltage curves of Sm-L-Au and Sm-D-Au NCs at the magnetic field directions of \textbf{+B} (red), \textbf{–B} (blue), and \textbf{0B} (black). We measured 180--650 current-voltage curves from more than 20 individual Au NCs and estimated the error bar by the standard deviation. (c) and (d) The MR of Sm-L-Au and Sm-D-Au NCs changing with bias under different external magnetic field directions. The Sm-L-Au and Sm-D-Au NCs were prepared by treating the L-Au and D-Au NCs with sodium sulfide, respectively. \textcolor{black}{The I-V characterizations are performed by cAFM.}}
    \label{figure4}
\end{figure}

The MR is positive for L/D-Au NCs as shown in Figure~\ref{figure2}. We discovered that the sign of MR is actually sensitive to the chemical composition of the surface molecular layer on chiral Au NCs. We further treated the chiral Au NCs with sodium sulfide, resulting in additional sulfide adsorption into the CTAB layers on the Au NCs~\cite{chen2024stabilizing}. For clarity, we designate the sulfide-modified L- and D-Au NCs as Sm-L-Au and Sm-D-Au NCs, respectively. X-ray photoelectron spectroscopy studies confirm the presence of sulfide on their surfaces (Extended Data Figure \ref{exFigure14}). SEM and spectra studies demonstrate that Sm-L-Au and Sm-D-Au NCs maintain similar chiral morphologies and optically chiral properties to their non-modified counterparts (Extended Data Figures \ref{exFigure12} and \ref{exFigure13}). Dynamic light scattering reveals slight size decreases for Sm-L-Au and Sm-D-Au NCs, attributed to the partial desorption of CTAB upon sulfide modification (Extended Data Figure \ref{exFigure15} and Extended Data Table \ref{exTable3}).

We investigated the MR of Sm-L-Au and Sm-D-Au NCs using the same cAFM setup (Figure~\ref{figure4} and Extended Data Figure~\ref{exFigure17}). At zero magnetic field, sulfide-modified NCs exhibit significantly higher conductance than pristine NCs, likely due to the thinner CTAB molecular layer. Despite this conductance change, the asymmetric nature of MR persists. For Sm-L(D)-Au NCs, \textbf{–B} (\textbf{+B}) field dramatically reduces the resistance by several times, while \textbf{+B} (\textbf{–B}) barely affects it. In contrast to L(D)-Au NCs, Sm-L(D) NCs display negative MR, as shown in Figure~\ref{figure4}. This sign change in MR suggests that the magnetotransport properties are intimately related to the characteristics of the molecular shell layer.

\begin{figure}
    \centering
    \includegraphics[width=0.8\linewidth]{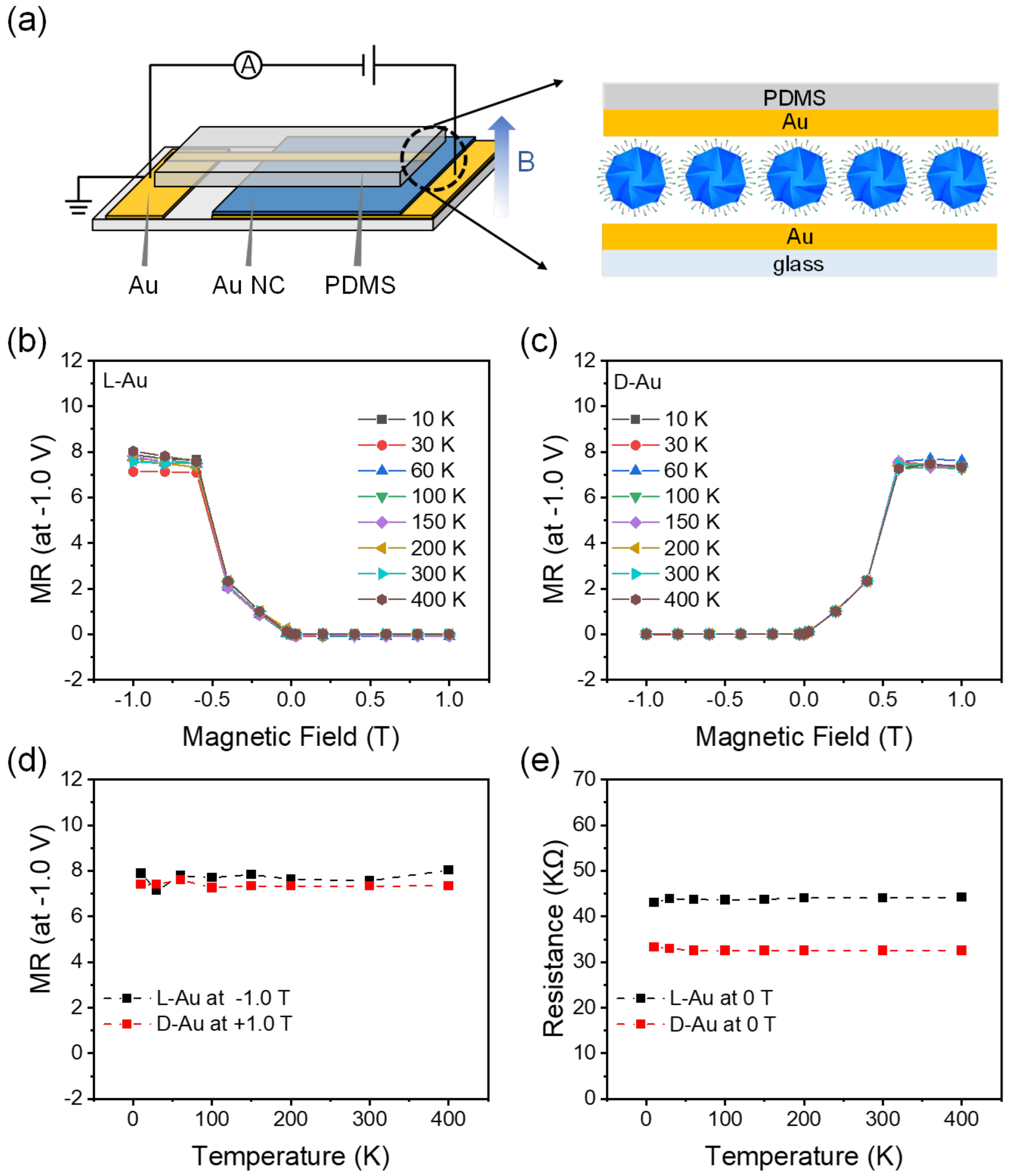}
    \caption{
    {\textbf{Magneto-transport and asymmetric MR for chiral Au NCs in solid-state device. }(a) Left: schematic illustration of the Au/chiral Au NC monolayer/Au solid-state device. Right: cross-sectional image of the Au/chiral Au NC monolayer/Au junction. (b) and (c) The MR of L-Au and D-Au NC monolayers measured under magnetic field ranging between -1.0 to 1.0 T and temperature between 10 and 400 K. (d) The trends of MR with temperature for L-Au NC monolayer at -1.0 T (black) and D-Au NC monolayer at +1.0 T (red). (e) The trends of resistance with temperature for L-Au NC monolayer (black) and D-Au NC monolayer (red) at 0 T. Both the MR and resistance are recorded at -1.0 V. }}
    \label{figure5}
\end{figure}

\section{
\textcolor{black}{
Magnetoresistance in solid-state devices}} \label{sec:device}

\textcolor{black}{For chiral molecules, cAFM measurements typically show high MR, whereas the corresponding devices exhibit significantly lower MR ($\sim1$\%)~\cite{Kiran2016,bloom2016spin,Al2018,Liu2020}. In contrast, we can reproduce the large MR in solid-state devices similar to the cAFM results for chiral Au NCs. These devices enable us to perform systematic measurements at varying temperature and magnetic field.
}

\color{black}{
The solid-state devices were constructed by preparing the Au NC assembled monolayers on a template-stripped Au electrode and placing a PDMS (polydimethylsiloxane)-embedded Au electrode on top of the Au NC monolayers (Figure~\ref{figure5}{a} and Extended Data Figure~\ref{exFigure28}). The I-V curves through the Au NC monolayer junctions could be recorded under external magnetic fields and given temperature (Extended Data Figures~\ref{exFigure30} and \ref{exFigure31}).} 
\color{black}{Similar to the results obtained by cAFM, the MR of chiral Au NC monolayers in solid-state devices exhibits a unidirectional response to the magnetic field direction. This MR response can be reversed by altering the chirality of Au NC and is independent of the applied order of the magnetic field (Figures~\ref{figure5}{b}, \ref{figure5}{c}, and Extended Data Figures ~\ref{exFigure24} and ~\ref{exFigure26}). The observed MR increases with magnetic field strength and saturates near 0.6 T for both L- and D-Au NC monolayers. 
The differences in MR saturation points between the solid-state device and cAFM measurements can be attributed to the number of chiral Au NCs involved and the types of electrodes used in these two methods. The cAFM measurement was conducted at single-particle level. In contrast, the contact area of the device is approximately 20,000 \textmu m$^2$, encompassing nearly one million Au NCs. A larger contact area introduces more defects, such as pin-hole or aggregation, causing the MR to reach saturation at higher magnetic fields. The different types of electrode contacts lead to various electrode-Au NC interfaces. 
As a control, the current of A-Au NC monolayer remains unchanged with the magnetic field direction even in the presence of a strong magnetic field (Extended Data Figures ~\ref{exFigure24} and ~\ref{exFigure27}).
More importantly, The current through both L- and D-Au NC monolayer junctions are temperature-independent, regardless of whether a magnetic field is applied. Correspondingly, the MR of L- and D-Au NC monolayers remains unaffected by temperature (Figures~\figref{figure5}{b}-\figref{figure5}{e}). These observation indicate a tunneling-dominated charge transport mechanism. Additional AC lock-in measurements of MR for both L- and D-Au NC monolayers corroborate the above results (Extended Data Figure~\ref{exFigure29}). 
}

\color{black}
\section{Discussions}

The MR observed in chiral NCs exhibits unique characteristics that distinguish it from known magneto-transport phenomena. Firstly, it is independent of current direction and violates the general Onsager's reciprocal relation. Secondly, its activation is triggered by a selected magnetic field direction, showing the asymmetric feature. These behaviors contrast sharply with established magneto-transport mechanisms:
(i) In nonmagnetic metals, the ordinary MR, induced by the Lorentz force, is proportional to $\mathbf{B}^2$ and symmetric for $\pm \mathbf{B}$~\cite{ashcroft1976solid}. (ii) In chiral conductors, EMCA produces an additional MR proportional to $\mathbf{B\cdot I}$, dependent on both current direction and amplitude~\cite{Rikken2001}. (iii) The CISS effect typically yields MR that is symmetric between opposite electrode magnetizations (\eg, Refs.~\cite{Xie2011,Liu2020,bloom2016spin,safari2024spin}) relative to the paramagnetic case.
Our observed MR deviates from these established phenomena. Unlike EMCA, but similar to CISS, it violates Onsager's relation, showing varied resistance at ±B in the zero-bias limit. However, it also differs from typical CISS MR in its asymmetry.
Existing theoretical models proposed to explain CISS MR~\cite{liu2021chirality,Yang2019b,Dubi2021,Fransson2022,hedegaard2023spin} fail to interpret our results. For instance, simple spin filter/polarizer models~\cite{liu2021chirality,Yang2019b,Wolf2022} only reproduce EMCA-like effects. More sophisticated models involving chirality-induced spin-transfer torques~\cite{Dubi2021,hedegaard2023spin} or vibration-enhanced spin-orbit coupling~\cite{Fransson2022,Ismael2021} require magnetic electrodes and predict symmetric MR for opposite magnetizations. These models cannot account for the pronounced MR asymmetry we observe using non-magnetic leads.
Given these discrepancies, we must conclude that our observed effect cannot be explained by ordinary MR, EMCA, or previously proposed CISS mechanisms. This necessitates the development of a new theoretical framework to understand the unique magneto-transport behavior in chiral NCs.

\begin{figure}
    \centering
    \includegraphics[width=0.9\linewidth]{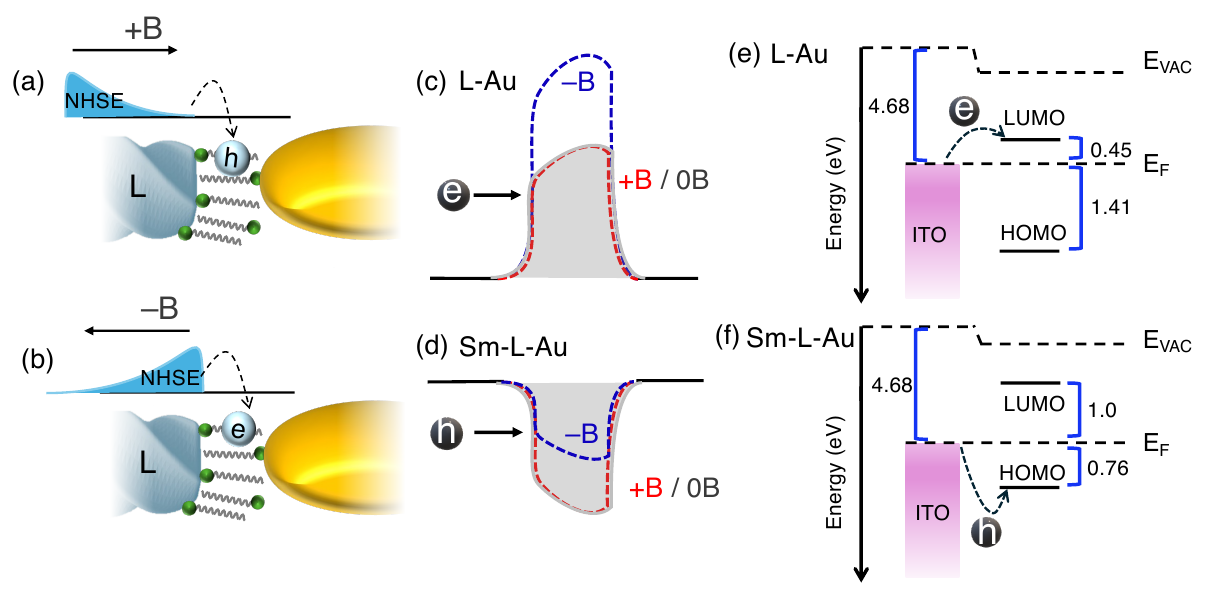}
    \caption{\color{black}{\textbf{The magnetic field modified tunneling barrier for chiral Au NCs.} (a)-(b) Schematics of the L-Au NC - molecular layer - electrode interface. In the presence of current-induced dissipation, the L-Au NC region exhibits the non-Hermitian skin effect (NHSE). Depending on the magnetic field (\textbf{B}) direction, wave functions are exponentially localized to (away from) NC - molecule interface, to pump electrons (holes) to the molecular layer. Reversing the chirality of Au NC will also flip the NHSE direction. (c) Because the molecular layer traps only electrons but not holes, the electron tunneling barrier through the molecular layer is increased at \textbf{–B} but unchanged at \textbf{+B} for L-Au NCs. (d) Similarly, the hole tunneling barrier is reduced at \textbf{–B} but unchanged at \textbf{+B} for Sm-L-Au NCs. (e)-(f) L-Au and Sm-L-Au NCs are electron and hole transport, respectively, according to their highest occupied molecular orbital (HOMO) and lowest unoccupied molecular orbital (LUMO) alignment with the electrode Fermi energy ($E_F$). The energy levels were measured by UPS and LEIPS. The work functions of ITO and Pt are measured with respect to the vacuum level ($E_{\mathrm{VAC}}$). As current flows, electron trapping will lift up both HOMO and LUMO, leading to the barrier change illustrated by (c)-(d).}}
    \label{figure6}
\end{figure}

\begin{table}
    \centering
    \caption{Fitted barrier heights ($U$) according to Eq.~\eqref{equation1} at different magnetic fields (\textbf{B}) for L/D and Sm-L/D-Au NCs from the experimental data in Figures~\ref{figure2} and \ref{figure4}.}
    \label{table1}
    \begin{tabular}{ccccc}
    \hline
         & L & D & Sm-L & Sm-D \\
         \hline
       \textbf{0B}  & 0.57 & 0.61 & 0.98 & 0.98 \\
       \textbf{+B}  & 0.58 & 1.15 & 0.97 & 0.56\\
       \textbf{-B}  & 1.26 & 0.58 & 0.54 & 0.93\\
       \hline
    \end{tabular}
\end{table}

To understand the unconventional MR, we first adopted a barrier-modulation model to analyze the observed asymmetric MR. As mentioned above, the insulating molecular shell covered on the Au NC determines the main resistance. The molecular CTAB layer is $\sim$3 nm thick~\cite{mosquera2023surfactant} and appears between the chiral Au NC and both electrodes including cAFM tip (Pt) and ITO. We approximate the molecular layer as an electron/hole tunneling barrier and attribute the resistance change caused by the magnetic field to the tunneling barrier modification. In the following, we employ a simplified one-dimensional barrier model and extract the barrier information by fitting the experimental I-V curves. Thereafter, we discuss the mechanism of barrier modification related to chirality and magnetic field. 

The Simmon’s model~\cite{Simmons1963} was modified to simulate a barrier tunneling problem,
\begin{equation} \label{equation1}
 I=g[(U+eV/2) e^{-A\sqrt{U+eV/2}}-(U-eV/2) e^{-A\sqrt{U-eV/2}} ](1+\gamma V)
\end{equation}
where $U$ is the barrier height, $A$ is a parameter related to the effective mass and barrier width, $g$ is a parameter in the unit of conductance, and $\gamma$ characterizes the anisotropy between $\pm V$ (\textit{e.g.}, caused by the device asymmetry). We use Eq.~\eqref{equation1} to fit our I-V data for a given magnetic field and extract $U,A,g,\gamma$. In principle, applying \textbf{B} only modifies the barrier $U$ but keeps other parameters unchanged. Thus, we use the same $A,g,\gamma$ but different $U$ in the fitting for I-V curves at $\mathbf{\pm B}$ and \textbf{B} = 0 from the same Au NC. The fitted I-V curves were shown in Figures \ref{figure2} and \ref{figure4}.
We illustrate the barrier modulation and summarize the fitting barrier in Table \ref{table1} (see details in Extended Data Table. \ref{exTable4}). One can find that independent fittings for L- and D-Au NCs give very similar barrier height ($U\approx 0.6$ eV) at zero field. The \textbf{–B (+B)} field enhances the barrier significantly to about 1.2 eV for L(D)-Au NCs. In contrast, \textbf{–B (+B)} reduces the barrier from about 1 eV to around 0.5 eV for Sm-L(D)-Au NCs. 

\textcolor{black}{According to the barrier-modulation model, MR can be regulated by altering the tunneling barrier of molecular shell. To verify the role of tunneling barrier in asymmetric MR, C$_n$TAB with varying chain lengths (n = 16, 12, 8, and 4) were applied to adjust the molecular shell barrier. The I-V curves of L-Au NCs with different C$_n$TAB molecular shells were measured at \textbf{–B} and \textbf{0B} (see Extended Data Figure \ref{exFigure23}).  All L-Au NCs showed the same trend that the current at \textbf{–B} is lower than that at \textbf{0B}. Moreover, under both \textbf{–B} and \textbf{0B} magnetic fields, the current increased approximately exponentially as the molecular chain length decreased. Correspondingly, the MR of L-Au NCs dropped from 4800\% to 300\% as the chain length from C$_{16}$TAB to C$_4$TAB. These results clearly demonstrate the dependence of MR on the thickness of molecular shells.}

To understand the opposite trend of barrier modification between L/D- and Sm-L/D-Au NCs, we performed further experiments and revealed that L/D- and Sm-L/D-Au NCs exhibit opposite charge carriers with electrons and holes, respectively, in transport. Considering the whole NC as a single particle, we measured the highest occupied molecular orbital (HOMO) and the lowest unoccupied molecular orbital (LUMO) using ultraviolet photoelectron spectroscopy (UPS) and low energy inverse photoemission spectroscopy (LEIPS), respectively, at \textbf{B} = 0 (see Figures \figref{figure6}{e} and \figref{figure6}{f} and Extended Data Figures \ref{exFigure18} and \ref{exFigure19}). For L/D-Au NCs, the LUMO energy is much closer to the Fermi energies of ITO compared to the HOMO energy. This indicates that electron rather than hole tunneling is dominant in the charge transport. Here, we can approximate the LUMO offset to electrode Fermi energy as the effective tunneling barrier, \textcolor{black} { $0.4\sim 0.5$ eV as shown in Figure~\figref{figure6}{e}.} This barrier is smaller than the fitted value from the simplified model but still in the same order of magnitude.
Because of different work functions between Pt (4.93 eV) and ITO (4.68 eV), the barrier profile (at zero bias) will be slightly tilted at zero bias as illustrated in Figure ~\figref{figure6}{c}.  
For Sm-L/D-Au NCs, however, the HOMO level is closer to the electrode Fermi energies than LUMO, indicating the hole transport nature. The effective barrier for holes is \textcolor{black} { $0.7\sim0.8$ eV here.} 

Known opposite tunneling carriers between L/D Au-NCs and Sm-L/D-Au NCs, we can rationalize their opposite barrier modulations by one scenario. Both the increase of electron barrier (L/D) and decrease of hole barrier (Sm-L/D) point to an up-shift of the NC electronic potential. In experiments, chiral Au NCs are all positively charged before and after sulfide modification, as revealed by the positive Zeta potential in Extended Data Figure~\ref{exFigure16} and Extended Data Table \ref{exTable3}. Thus, their molecular layer prefer to trap extra electrons rather than holes. Then, electron-trapping increases both HOMO and LUMO energies of chiral NCs upon applying a selected magnetic field.

\textcolor{black}{
Our observation coincides with the magnetochiral charge pumping mechanism proposed to understand CISS recently~\cite{Zhao2025}.  
According to this theory, a non-Hermitian skin effect (NHSE)~\cite{Yao2018,OkumaSatoReview2023} exists in chiral Au NCs due to coexisting chirality, magnetic field, and current-induced dissipation. Here, NHSE wave functions are exponentially localized toward to or away from the NC - molecule interface, pumping electrons or holes to the molecular layer which depends on the magnetic field direction and chirality. 
For example, electrons (holes) are pumped at \textbf{–B} (\textbf{ +B}) for L-Au and Sm-L-Au NCs, as illustrated in Figure~\ref{figure6}. {Because the capping molecular layer only traps electrons, the electron potential is increases at \textbf{–B} but unchanged at \textbf{+B}. Consequently, the electron barrier is increased for L-Au NCs while the hole barrier is decreased for Sm-L-Au NCs.}
Because of the selective charge trapping, the tunneling resistance is dramatically modified by one magnetic field but remains unchanged in the opposite field, explaining the strong asymmetry in MR. 
Here, switching chirality exhibits a similar resistance response to flipping the magnetic field. 
In this scenario, {NHSE} controls the charge pumping by magnetic field and chirality and the eventual charge trapping changes the resistance, which is dubbed magnetochiral charge pumping~\cite{Zhao2025}. {NHSE} appears only in the nonequilibrium phase when current flows while charge trapping persists even if current diminishes. Thus, the zero-bias resistance changes dramatically upon flipping magnetic field because of the non-adiabatic charge trapping, circumventing the constraint of Onsager's relation.
}

\textcolor{black}{
The magnetochiral charge pumping, which involves NHSE and charge trapping, can unify different chirality-driven MR effects including the asymmetry MR, CISS MR and EMCA in one framework. 
Besides dissipation, NHSE requires breaking both time-reversal symmetry and inversion symmetry in a lattice. (i) In chiral Au NCs, the inversion symmetry is broken by chirality and the time-reversal symmetry is broken by the magnetic field. Here, conducting electrons from the Au core feel \textbf{B} sensitively via the Lorentz force, instead of the negligible Zeeman effect, and show cyclotron motions (the orbital effect). When current flows, electrons move in a spiral pathway that can coincide or contradict the NC chiral morphology, leading to different localization pattern of the wave function along the current or \textbf{B}, i.e., {NHSE}. 
We note that the magnetic length is about $l_B=25.6 ~\mathrm{nm} /\sqrt{B}=148$ nm for B = 0.03 T, supporting the cyclotron motion picture. 
(ii) In chiral conductors, only NHSE leads to EMCA~\cite{Yi2020} because of missing charge trapping. 
(iii) In CISS with ferromagnetic electrode, electrons at the interface can simultaneously feel chirality and exchange splitting and thus exhibit NHSE at the interface and charge trapping in the insulating molecule region. Without the ferromagnetic contact, chiral molecules cannot feel the magnetic field because of missing free electrons and then exhibit no NHSE, leading to vanishing MR.}

\section{Perspectives and Conclusions}

Our work opens up exciting future directions for enantiomer-selective modification of surface potentials on chiral NCs using external magnetic fields, with potential applications in controlling chemical and biological reactions. For instance, a negative magnetic field increases the electron potential energy for L-/Sm-L-Au NCs while leaving that of D-/Sm-D-Au NCs unchanged. This feature could enable fine-tuned regulation of surface reactions, such as redox processes and electrostatic interactions, via external magnetic fields. Such control could lead to the development of novel magnetically driven strategies for metal-catalyzed reactions \cite{zuo2023mechano} and specific enzymatic binding interactions.
Furthermore, the magneto-enantiomer selection of chiral NCs demonstrates sensitivity to surface ligands, as evidenced by the opposite MR signs displayed by L-Au and Sm-L-Au NCs under the same magnetic field. This property suggests potential applications of these chiral NCs as a new class of indicators for distinguishing molecules with different binding affinities to chiral surfaces.
Additionally, the unique responses of chiral NCs to magnetic fields could be harnessed for the controllable separation of molecular mixtures. By applying a designated magnetic field, it may be possible to selectively modify the surface properties of specific enantiomers, facilitating their separation from a mixture.
These potential applications highlight the broad implications of our findings, spanning from fundamental surface chemistry to practical applications in catalysis, molecular recognition, and enantiomer separation. Future research in this direction could lead to innovative technologies leveraging the interplay between chirality, magnetism, and surface chemistry at the nanoscale.

In summary, we observed remarkable chirality-driven asymmetric MR in chiral Au NCs at the single nanoparticle level. This unusual MR exhibits distinct differences from CISS and EMCA, characterized by its asymmetric nature and absence of ferromagnetic electrodes. Notably, a simple sulfide modification of the chiral Au NCs enables switched positive and negative MR responses. We attribute the salient MR behavior to the asymmetric charge trapping influenced by both the magnetic field and chirality.
Our work establishes the theory framework to understand different chirality-driven phenomena and paves a path to perform more quantitative calculations in the future. The implications of this unique MR mechanism extend beyond CISS and hold potential in the application of advanced crystalline electronic/spintronic devices and single-particle sensors. Our work indicates an exotic surface potential activation of chiral Au NCs and remote control of charge transfer in chiral Au NCs by magnetic fields. The charge transfer under a magnetic field can be selectively inhibited or enhanced by controlling their surface chemistry, providing a new strategy for manipulating enantiomer-specific chemical or biological reactions and the development of chemical sensors. 

\section*{Methods}

\textbf{Synthesis of CPC-capped Au seeds. }The CPC-capped Au seeds were synthesized according to procedures published elsewhere \cite{niu2009selective,Mirkin2014}, and they were used for the synthesis of different types of Au NCs. The major samples and their corresponding abbreviations in this report are listed in Extended Data Table~\ref{exTable1}.

\noindent \textbf{Synthesis of chiral Au NCs.} The synthesis of chiral Au NCs is realized through a Cys-directed growth method, which is based on the enantioselective interactions of L- or D-Cys with R- or S-$\{hkl\}$ chiral Au crystal facets ($h > k > l > 0$)~\cite{lee2018amino,wu2022synthesis}. The R- or S-$\{hkl\}$ chiral Au crystal facet is defined as the clockwise or anticlockwise rotational progression among the (111), (100), and (110) microfacets around the kink atoms. In a typical synthesis, a growth solution was prepared by gently mixing 2.025 mL diluted water, 1.8 mL of 100 mM CPC and 100 mM CPB solution (CPC:CPB=1:3), 0.2 mL of 10 mM HAuCl$_4$ solution, 1.4 mL of 0.1 M \textcolor{black}{ascorbic acid (AA)} solution, 12.5 $\mu$L of 0.1 mM L- or D-Cys solution at 30 $^\circ$C. With the addition of Au seed solution, the growth of chiral NCs was started and continued at 30 $^\circ$C for 2 h. Then, the solution was centrifuged (6000 rpm, 3 min) and washed twice and then stored in 1 mM CTAB solution. 

For the addition of 3, 6, 12, 24, 48, and 80 $\mu$L of seeds, the final sizes of the chiral Au NCs are 475.5±6.4 nm, 377.2±6.3 nm, 350.5±5.4 nm, 225.7±5.6 nm, 188.8±6.2, and 150.2±4.8 nm, respectively. 

The CATB bilayer structure involves the adsorption of bromide anions on the Au surface, establishing an electrostatic interaction with the $\mathrm{(CH_3)_3N^+}$ head group of the cetyltrimethylammonium cation ($\mathrm{CTA^+}$)~\cite{mosquera2023surfactant}. This arrangement leads to the outward orientation of the long hydrophobic alkane tail of $\mathrm{CTA^+}$, promoting the formation of a second layer through interactions between the alkane chains of other CTA+ cations. Consequently, the cationic $\mathrm{(CH_3)_3N^+}$ groups remain on the external sides of the bilayer and the chiral Au NCs exhibit a positive charge.

\noindent \textbf{Synthesis of achiral Au NCs from racemic Cys.} Achiral Au NCs from racemic Cys were synthesized with the addition of 6.25 $\mu$L of 0.1 mM L-Cys and 6.25 $\mu$L of 0.1 mM D-Cys solution while keeping other parameters constant.

\noindent \textbf{Synthesis of achiral Au NCs in the absence of Cys.} A 0.2 mL of 10 mM HAuCl$_4$ solution, 40 $\mu$L of 100 mM AA solution, and 80 $\mu$L of CPC-capped Au seeds solution were added to 4 mL of 50 mM CTAC solution at 30 $^\circ$C in sequence. The mixture was thoroughly mixed and undisturbed for 30 min. Then, the solution was centrifuged (6000 rpm, 3 min), washed twice and redispersed in water.

\noindent \textbf{Modification of achiral Au NCs with L- or D-Cys.} 80 $\mu$L of 100 mM L- or D-Cys solution was added into 2 mL of 4 mM the achiral Au NC solution (synthesized in the absence of Cys). The mixture was thoroughly mixed and incubated for 1 h at 30 °C. Then, the solution was centrifuged (6000 rpm, 3 min), washed once, and stored in 1 mM CTAB solution.

\noindent \textbf{Modification of chiral Au NCs with sulfide.} 1 mL of 2 mM $\text{Na}_2$S solution was added into 2 mL of the L- or D-Au NC solution (1 mM). The mixture was thoroughly mixed and incubated for 15 min at 30 °C. Then, the solution was centrifuged (6000 rpm, 3 min), washed once, and stored in 1 mM CTAB solution.

\noindent \textcolor{black}{
\textbf{Modification of L-Au NCs with $\text{C}_n$TAB molecular shells of different chain lengths.}
The $\text{C}_n$TAB molecular shells of varying chain lengths (n = 12, 8, 4) were employed as ligands for L-Au NCs. Typically, the synthesized L-Au NCs were centrifuged (6000 rpm, 3 min) and washed with $\text{C}_{12}$TAB, $\text{C}_8$TAB, or $\text{C}_4$TAB solution, then stored in 1 mM $\text{C}_{12}$TAB, $\text{C}_8$TAB, or $\text{C}_4$AB solution, respectively.
}

\noindent \textbf{Characterization.} Scanning electron microscopy (SEM) images were taken using a ZEISS Sigma 300 field-emission microscope equipped with an accelerating voltage of 20 kV. Extinction and circular dichroism (CD) spectra were obtained using a J-820 spectropolarimeter instrument (JASCO). Transmission electron microscope (TEM) were obtained from a Japan's Hitachi H-600 TEM operating at 100 kV. Extinction spectra were collected by Shimadzu UV-1800 UV-vis spectrophotometer. X-ray photoelectron spectroscopy (XPS) spectra were performed with an ESCALAB MKII (VG Co., UK) spectrometer with an AlKa excitation source. Dynamic light scattering and zeta potential measurements were carried out on Malvern Zetasizer (Malvern Instrument Ltd, UK).  

\noindent \textcolor{black}{
\textbf{superconducting quantum interference device (SQUID) magnetometry.}
The magnetic moment (M) of Au NC powder samples was measured using an MPMS3 SQUID magnetic property measurement system (Quantum Design). At 300 K, the M-H curves were obtained in vibrating sample magnetometer (VSM) mode under an external magnetic field ranging from +0.5 T to -0.5 T.
To correct the background diamagnetic signal, M-H data at high magnetic field (\( >0.2 \,  \text{ T} \)) were extracted for linear fitting, the slope of which provides the diamagnetic susceptibility of the background and is subtracted from the M-H curve.}

\noindent \textbf{Preparation of individual Au NCs on the ITO. }The indium tin oxide (ITO) substrates ($R = 10\; \Omega$) were cleaned by sonication in ultrapure water ($18.2 \; M\Omega\cdot \text{cm}$) and ethanol for 10 min, respectively. Thereafter, the ITO were washed with ultrapure water and dried with nitrogen. The Au NC solution (50 $\mu$L) was dropped on the ITO conductive surface. After natural drying, the Au NCs were dispersed on ITO substrates.

\noindent \textbf{cAFM measurements. }The images of Au NC on ITO substrates were measured using a NX 10 Complete AFM System (Park, Korea) with non-contact mode (NCM). The current-bias (I-V) curves of single Au NC were recorded in conducting atomic force microscopy (C-AFM) mode by applying the bias (from -1.0 V to 1.0 V) to the single Au NC at loading force of 10 nN. The Pt coated tip (HQ: XSC111, spring constant: $\sim$ 42 N/m) were used, and the I-V curves via the ITO/single Au NC/Pt junction were collected under external magnetic fields (300 Oe). \textcolor{black}{The order of changing the external magnetic field direction was random, and the cAFM setup was reset when the magnetic field direction was altered.} For each type of Au NC, hundreds of I-V curves were collected from at least \textcolor{black}{20 $\sim$ 30} individual Au NCs with more than three parallel samples. The log$\left| I \right|$ values at -1.0 V were extracted for statistics. All statistical histograms exhibited a normal distribution. The AFM images before and after I-V measurements display the same Au NCs morphologies, indicating the intact Au NCs during the I-V measurements. \textcolor{black}{The position of the Au NCs would not change when applying the magnetic field. Because the magnetic susceptibility shows no chirality dependence (Extended Data Figure~\figref{exFigure25}), the field-induced motion cannot generate chirality-dependent resistance even if it exists.} 

For the measurements with varied loading forces, a force of 130 nN would cause damage to the sample, so the maximum force was limited to 100 nN. Various forces were applied in situ on the same Au NC, and 20-40 I-V curves were collected for each loading force. Under each magnetic field condition, 3-5 individual Au NCs were measured. While the absolute current values are slightly different among different Au NCs, the trends of current behavior and MR remain consistent.

\noindent \textcolor{black}{
\textbf{Fabrication of Au NC assembled monolayers.} The L-Au, D-Au, and A-Au NC assembled monolayers were prepared using a previously reported liquid/liquid interface self-assembly method, separately~\cite{Jin2019iscience}. In this process, 3 mL of Au NCs (4 mM) was added to a plastic container, followed by the addition of 460 $\mu$L of n-hexane to form a two-phase interface. Next, 3.7 mL of methanol was rapidly poured into the mixture to capture the nanoparticles at the hexane/water interface. Upon evaporation of the hexane, the nanoparticles self-assembled into a monolayer at the water/hexane interface. Finally, the nanofilms were carefully transferred from the air-water interface onto the template-stripped Au electrodes or hydrophilic Si substrates for the next characterizations.}

\noindent \textcolor{black}{
\textbf{Solid-state device.}
The assembled monolayers of chiral Au NCs were prepared on the template-stripped Au electrode. An Au electrode embedded in PDMS was then placed on top of the Au NC monolayer as the top electrode, constructing a chiral Au NC monolayer solid-state device for electric measurements. The temperature and magnetic field conditions were precisely controlled using a Lakeshore probe station, while current measurements were performed using a Keysight 1500 instrument.}

\noindent \textcolor{black}{
\textbf{AC lock-in measurements.}
The AC lock-in measurements were performed on the solid-state device of chiral Au NC monolayer junctions with temperature and magnetic field control provided by the Lakeshore probe station. The measurements were conducted using the Lakeshore M81-SSM instrument, with a VS-10 source module supplying the voltage source and a CM-10 source module handling current measurement. During the experiment, a -1.0 V DC bias combined with a 3 mV AC signal at a frequency of 177.1 Hz was used as the voltage source for the device. The current lock-in response at 177.1 Hz was recorded for the following MR calculations.}

\noindent \textbf{Energy levels measurements. }
The HOMO and LUMO of Au NCs thin film on ITO substrate were obtained by ultraviolet photoelectron spectroscopy (UPS) and low energy inverse photoemission spectroscopy (LEIPS), respectively. UPS measurements were carried out using an ESCALAB 250Xi (Thermo Fisher, USA) spectrometer with He I line (21.22 eV) at a bias of -5.0 V. LEIPS measurements were performed on a customized ULVAC-PHI LEIPS instrument with Bremsstrahlung isochromatic mode.

\section*{Acknowledgements}
This work is supported by the National Natural Science Foundation of China (Nos. 22374144 to W.N., 22374109 and 21974102 to G.C., 22072144 to F.W., 22102171 to Y.T., 22373026 to Z.X.), the Israel Science Foundation (ISF, No. 2932/21 to B.Y.), the Minerva-Weizmann project (B.Y.). National Key R\&D Program of China (2023YFA1407100 to Z.X.), and Guangdong Science and Technology Department (2021B0301030005, STKJ2023072, and GDZX2304005 to Z.X.). The authors thank Dr. Hui-Ying Sun from the Core Facility of Wuhan University for the magnetic measurement, Dr. Ye Zou from the Institute of Chemistry for LEIPS measurements, and Profs. Jun Zhang and Ronghua Liu from Wuhan University and Nanjing University for insightful discussions on SQUID characterizations.

\section*{Competing Interests} The authors declare no competing interests.\

\section*{References}
%\bibliography{references-chiral}
%\bibliographystyle{naturemag}

\clearpage
\setcounter{figure}{0}
\setcounter{table}{0}
\renewcommand{\figurename}{\textbf{Extended Data Fig.}}
\renewcommand{\tablename}{\textbf{Extended Data Table}}

\begin{table}
    \centering
    \caption{A list of sample names of Au NCs and their corresponding abbreviations.}
    \label{exTable1}
    \begin{tabular}{ccc}
    \hline
       No.  & Sample Name  & Abbreviation \\
       \hline
1 &	L-chiral Au NCs	&L-Au \\
2 &	D-chiral Au NCs	&D-Au\\
3&	Achiral Au NCs	&A-Au\\
4&	L-cysteine-modified achiral Au NCs	&Lm-A-Au\\
5&	D-cysteine-modified achiral Au NCs	&Dm-A-Au\\
6&	Sulfide-modified L-chiral Au NCs	&Sm-L-Au\\
7&	Sulfide-modified D-chiral Au NCs	&Sm-D-Au\\
        \hline
    \end{tabular}
\end{table}

\begin{figure}
\centering
\includegraphics[width=0.8\linewidth]{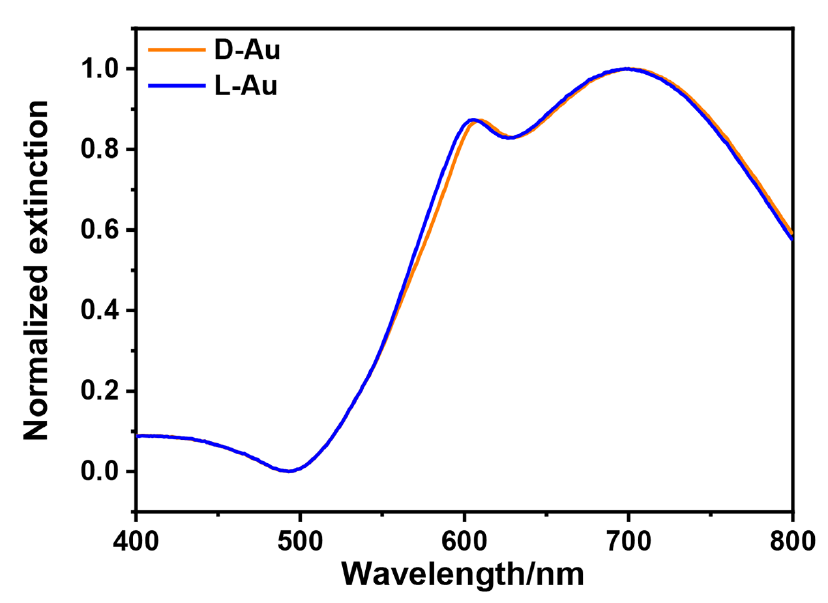}
\caption{The normalized extinction spectra of the chiral Au NCs shown in Figure~\ref{figure1}.}
\label{exFigure1}
\end{figure}

\begin{figure}[tp!]
\centering
\includegraphics[width=0.8\linewidth]{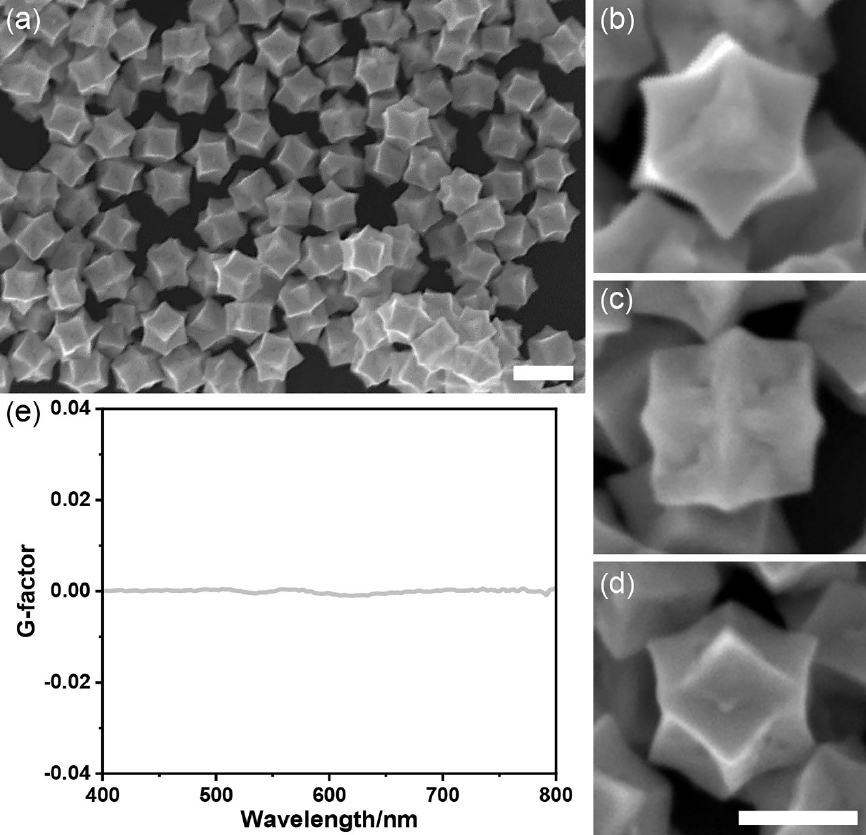}
\caption{SEM images of (a) achiral Au NCs synthesized with racemic Cys (A-Au), (b-d) SEM images viewed from the
$\braket{100}$,$\braket{111}$, and $\braket{110}$ axes of achiral Au NCs, respectively. (e) The g-factor spectra of achiral Au NCs. The size of A-Au is 151.1$\pm$4.5 nm. Scale bars: 200 nm in (a), 100 nm in (b-d).}
\label{exFigure2}
\end{figure}

\begin{figure}[tp!]
\centering
\includegraphics[width=0.8\linewidth]{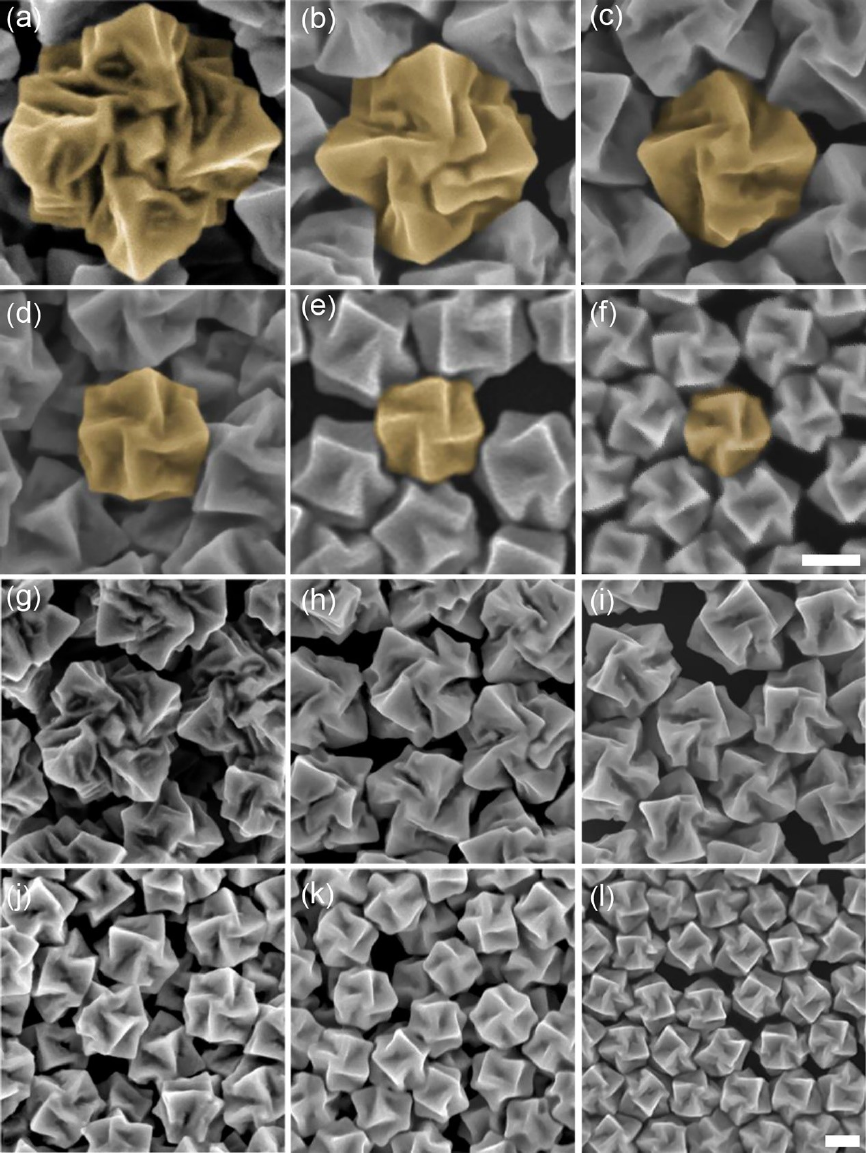}
\caption{SEM images of chiral Au NCs of different sizes synthesized in the presence of different volumes of Au seed solution (a, g) 3 µL, (b, h) 6 µL, (c, i) 12 µL, (d, j) 24 µL, (e, k) 48 µL, (f, l) 80 µL. L-Cys was used as the chiral inducer. Scale bars: 100 nm in (f and l). (a-e) share the same scale bar in (f). (g-k) share the same scale bar in (l). L-Au NCs are taken as examples.  }
\label{exFigure3}
\end{figure}

\begin{figure}[tp!]
\centering
\includegraphics[width=0.9\linewidth]{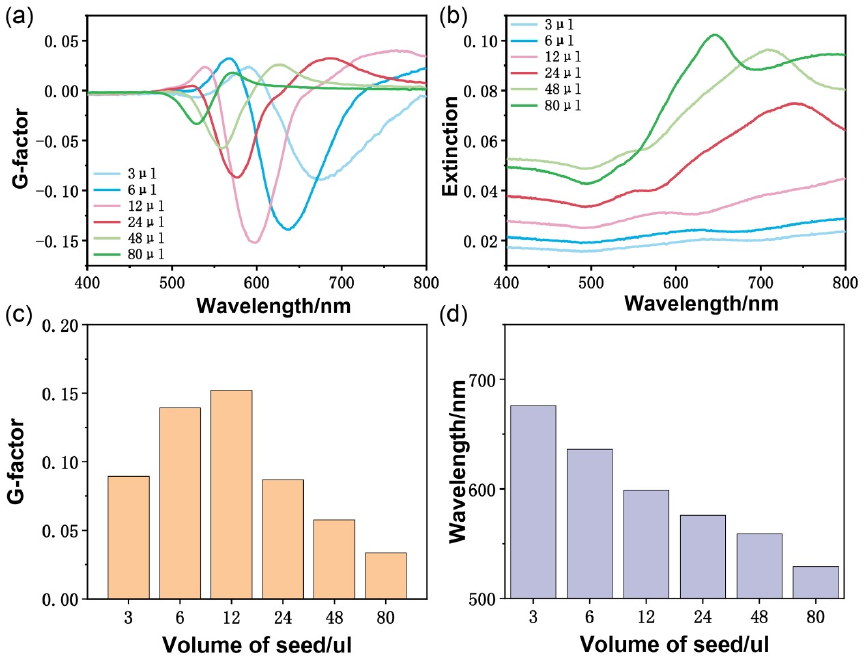}
\caption{(a) The g-factor spectra and extinction spectra of chiral Au NCs with different volumes of Au seed solution. (c) Peak value (absolute value) and (d) peak wavelength of g-factors spectra of the synthesized chiral nanoparticles different volumes of Au seed solution. L-Au NCs are taken as examples. }
\label{exFigure4}
\end{figure}

\begin{table}
    \centering
    \caption{A summary of chiral Au NCs synthesized with different volumes of seeds, their sizes, and their corresponding g-factors (L-Au NCs are taken as examples).}
    \label{exTable2}
    \begin{tabular}{ccc}
     \hline
       Seeds ($\mu L$)  & Size (nm)  & g-factor \\
       \hline
        3 & 475.5±6.4  & -0.088\\
        6 & 377.2±6.3  & -0.137\\
        12 & 350.5±5.4  & -0.147\\
        24 & 225.7±5.6  & -0.086\\
        48 & 188.8±6.2  & -0.051\\
        80 & 150.2±4.8  & -0.033\\
        \hline
    \end{tabular}
\end{table}

\begin{figure}[tp!]
\centering
\includegraphics[width=1\linewidth]{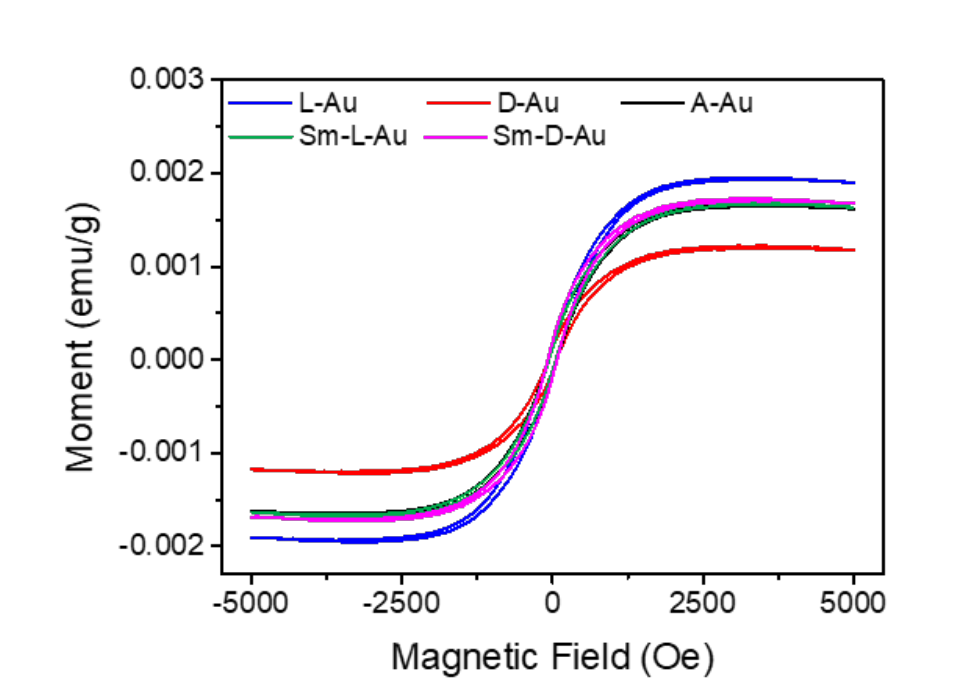}
\caption{\textcolor{black}{Magnetization curves for L-Au, D-Au, A-Au, Sm-L-Au, and Sm-D-Au NCs at 300 K.
Au NCs exhibit a very weak ferromagnetic behavior at 300 K, which is in line with the weak ferromagnetism of gold nanoparticles and nanocrystals reported in literature. We note that this behavior is similar among L, D, achiral and Sm-L/D NCs, which cannot account for the asymmetric, chirality-dependent and surface-sensitive MR in our experiments.
}}
\label{exFigure25}
\end{figure}

\begin{figure}[tp!]
\centering
\includegraphics[width=1\linewidth]{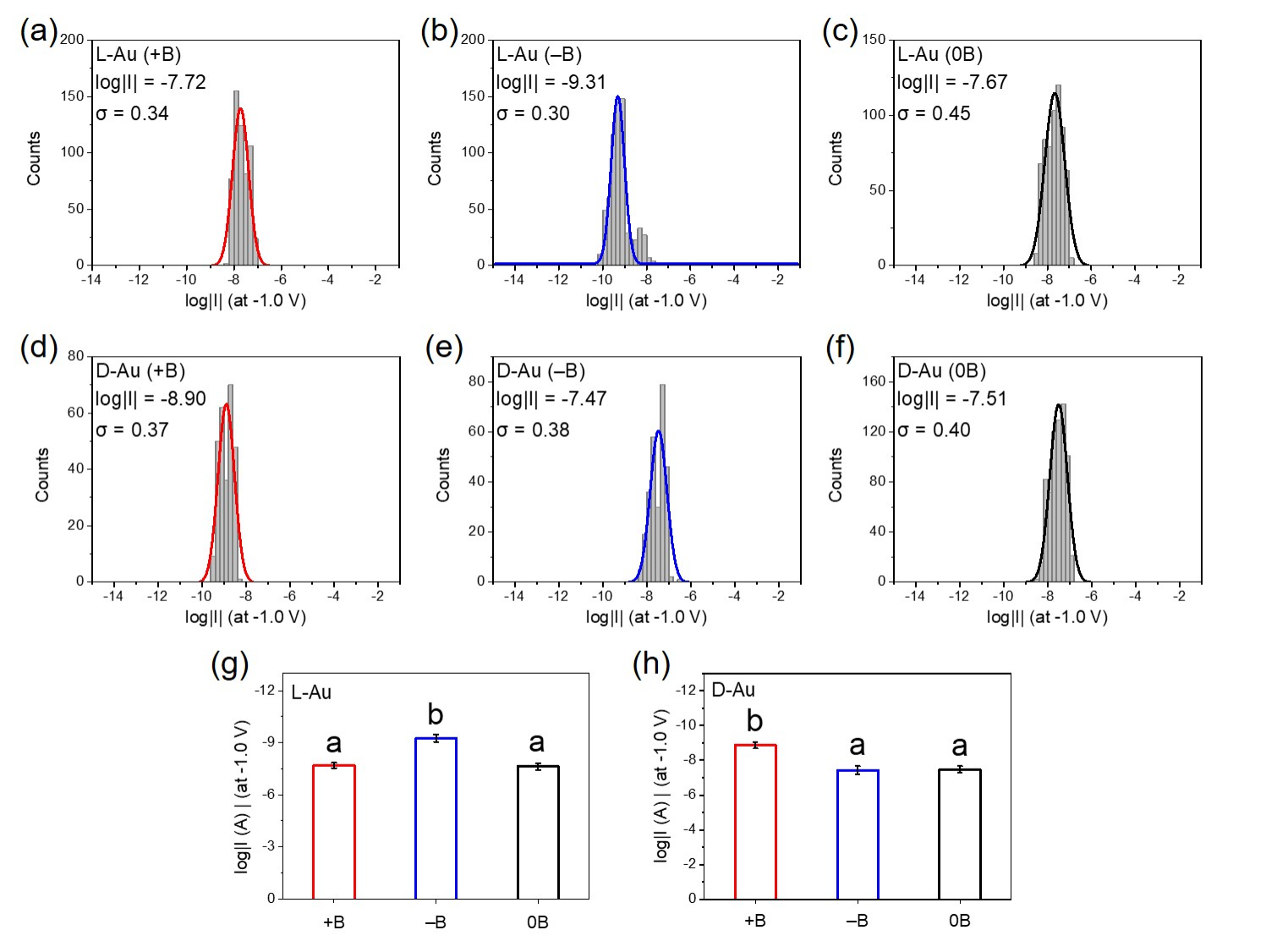}
\caption{Statistics of current ($log|I|$) for the L-Au NCs (a-c) and D-Au NCs (d-f) at the magnetic field directions of +\textbf{B}, –\textbf{B}, and 0\textbf{B}. The solid lines represent the Gaussian fits to these histograms. N = 320--679. (g) and (h) Corresponding $log|I|$ at -1.0 V with statistical significance for L-Au and D-Au NCs under magnetic field directions of +\text{B} (red), –\text{B} (blue), and 0\text{B} (black). Different letters represent significant differences at the P $<$ 0.05 level based on one-way analysis of variance (AVOVA). \textcolor{black}{The I-V characterizations are performed by cAFM.}}
\label{exFigure5}
\end{figure}

\begin{figure}[tp!]
\centering
\includegraphics[width=1\linewidth]{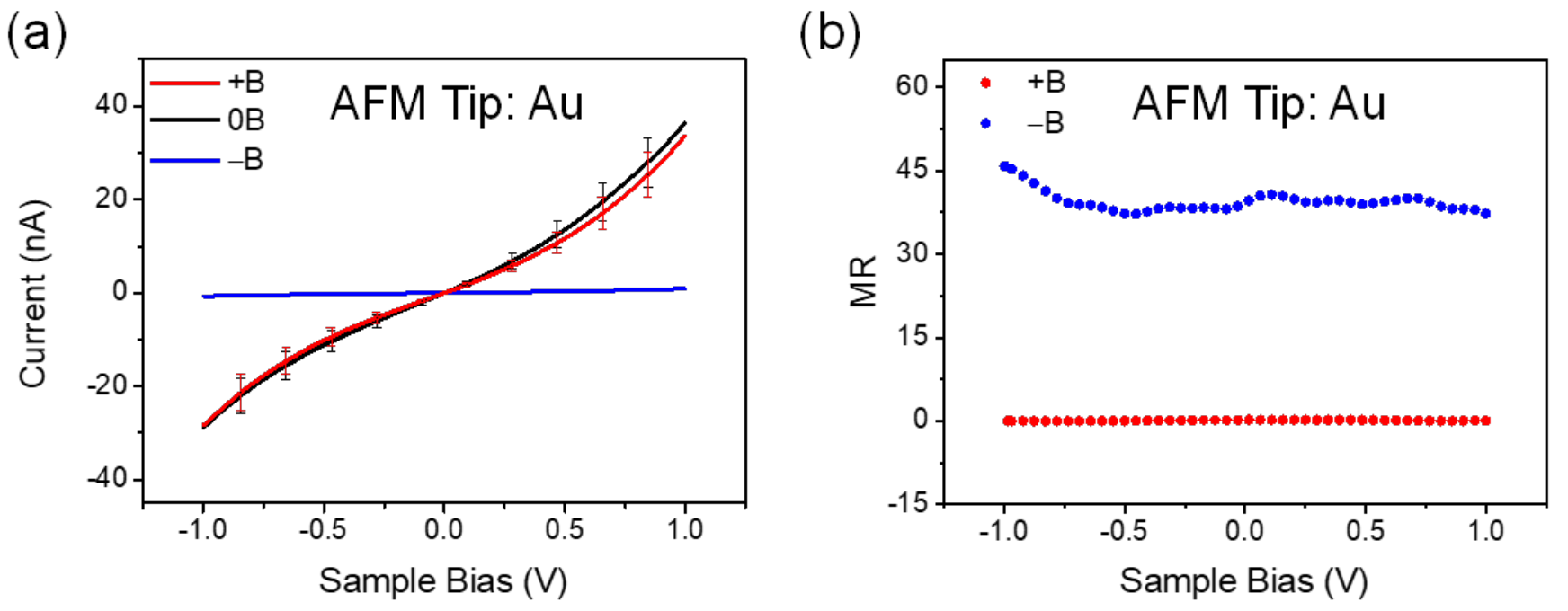}
\caption{\textcolor{black}{Magnetoresistance characterization of ITO/L-Au NC/Au junctions. (a) and (b) The experimental I-V curves and MR of L-Au at the magnetic field directions of \textbf{+B} (red), \textbf{-B} (blue), and \textbf{0B} (black). The I-V characterizations are performed by cAFM.}}
\label{exFigure21}
\end{figure}

\begin{figure}[tp!]
\centering
\includegraphics[width=0.5\linewidth]{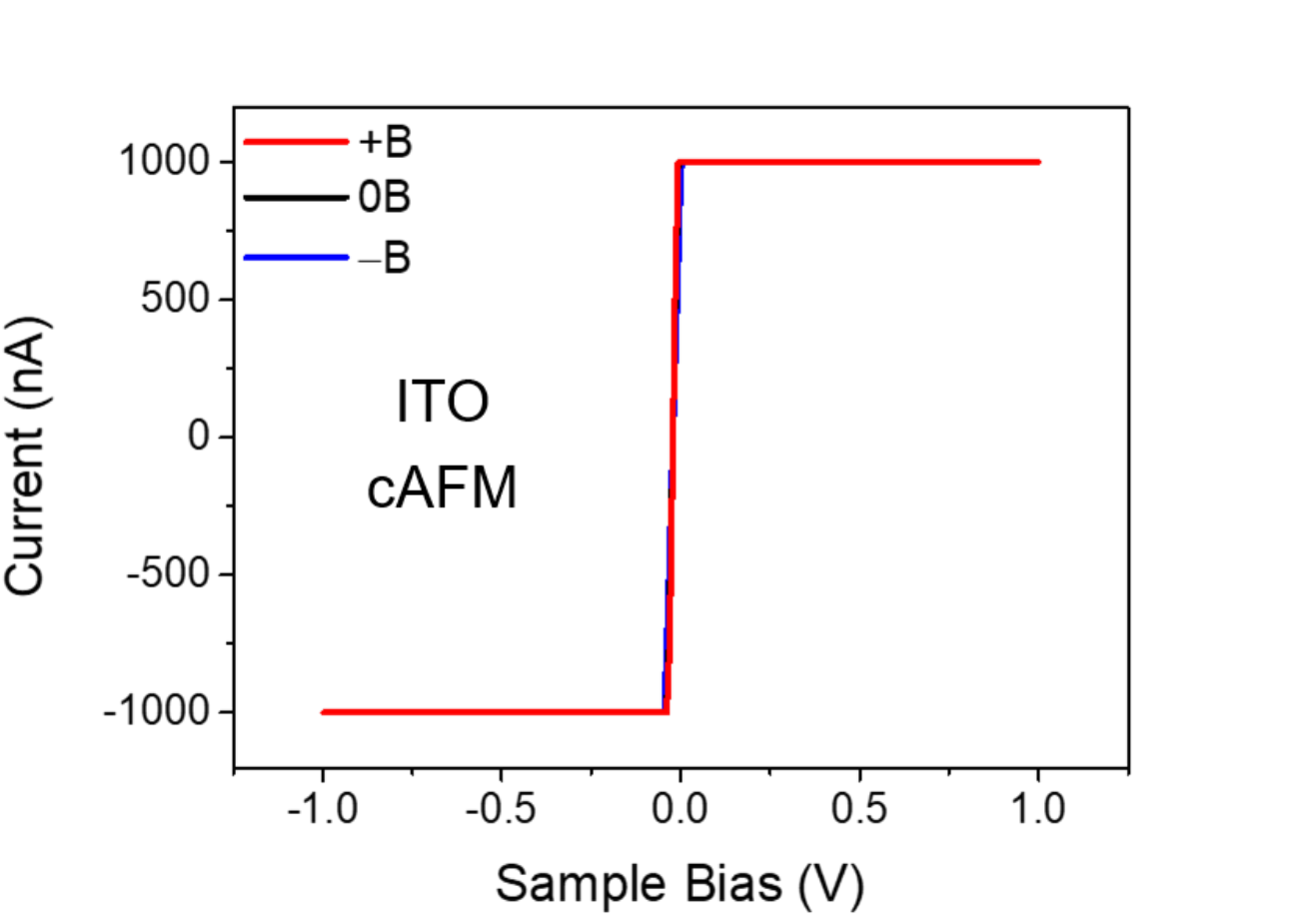}
\caption{\textcolor{black}{Magnetoresistance characterization of ITO under the magnetic field directions. The direct contact of ITO and Pt electrodes exhibits unchanged I-V curves under different magnetic field directions, with saturation occurring at $ \sim 50~\text{mV} $. The I-V characterizations are performed by cAFM.}}
\label{exFigure20}
\end{figure}

\begin{figure}[tp!]
\centering
\includegraphics[width=1\linewidth]{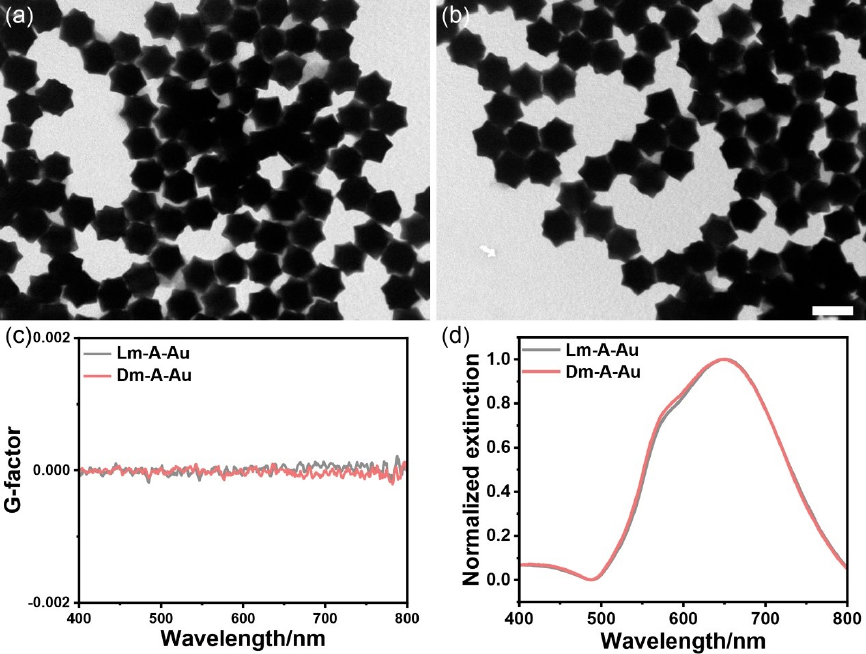}
\caption{TEM images of achiral Au NCs (155.5±5.6 nm in size) modified with L-Cys and D-Cys, respectively: (a) Lm-A-Au NCs, (b) Dm-A-Au NCs. (c) and (d) G-factor and normalized extinction spectra of the Lm-A-Au and Dm-A-Au NCs, respectively. Scale bar: 200 nm. (a) share the same scale bar in (b).}
\label{exFigure6}
\end{figure}

\begin{figure}[tp!]
\centering
\includegraphics[width=1\linewidth]{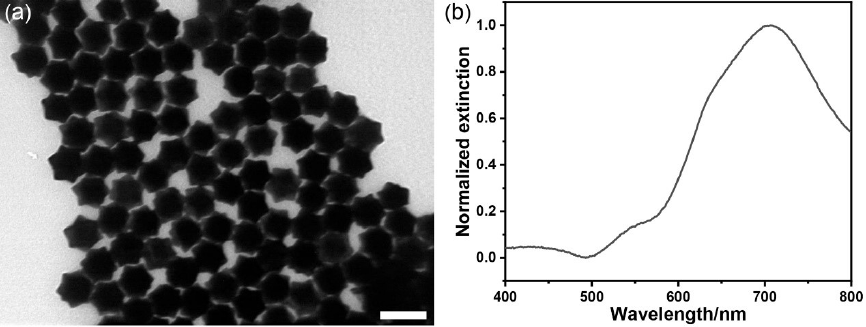}
\caption{(a) TEM image and (b) normalized extinction spectra of achiral Au NCs with the size of 155.5±5.6 nm synthesized in the absence of Cys. These achiral Au NCs were further used for the preparation of Lm-A-Au and Dm-A-Au NCs. Scale bar: 200 nm.}
\label{exFigure7}
\end{figure}

\begin{figure}[tp!]
\centering
\includegraphics[width=1\linewidth]{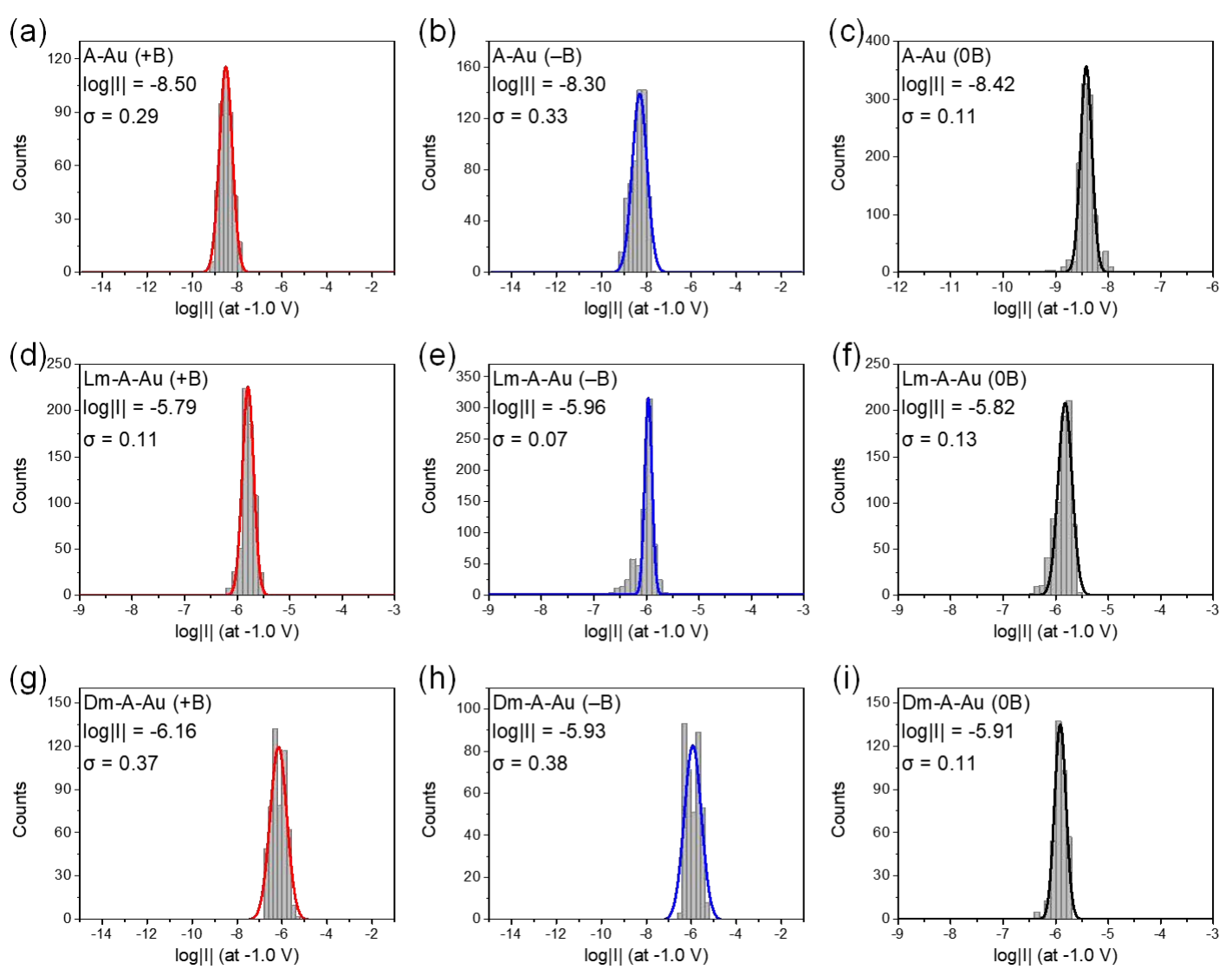}
\caption{Histograms of $log|I|$ with a Gaussian fit for the A-Au NCs (a-c), Lm-A-Au NCs (d-f), and Dm-A-Au NCs (g-i) under different directions of magnetic fields. N = 430-740. \textcolor{black}{The I-V characterizations are performed by cAFM.}}
\label{exFigure8}
\end{figure}

\begin{figure}[tp!]
\centering
\includegraphics[width=1\linewidth]{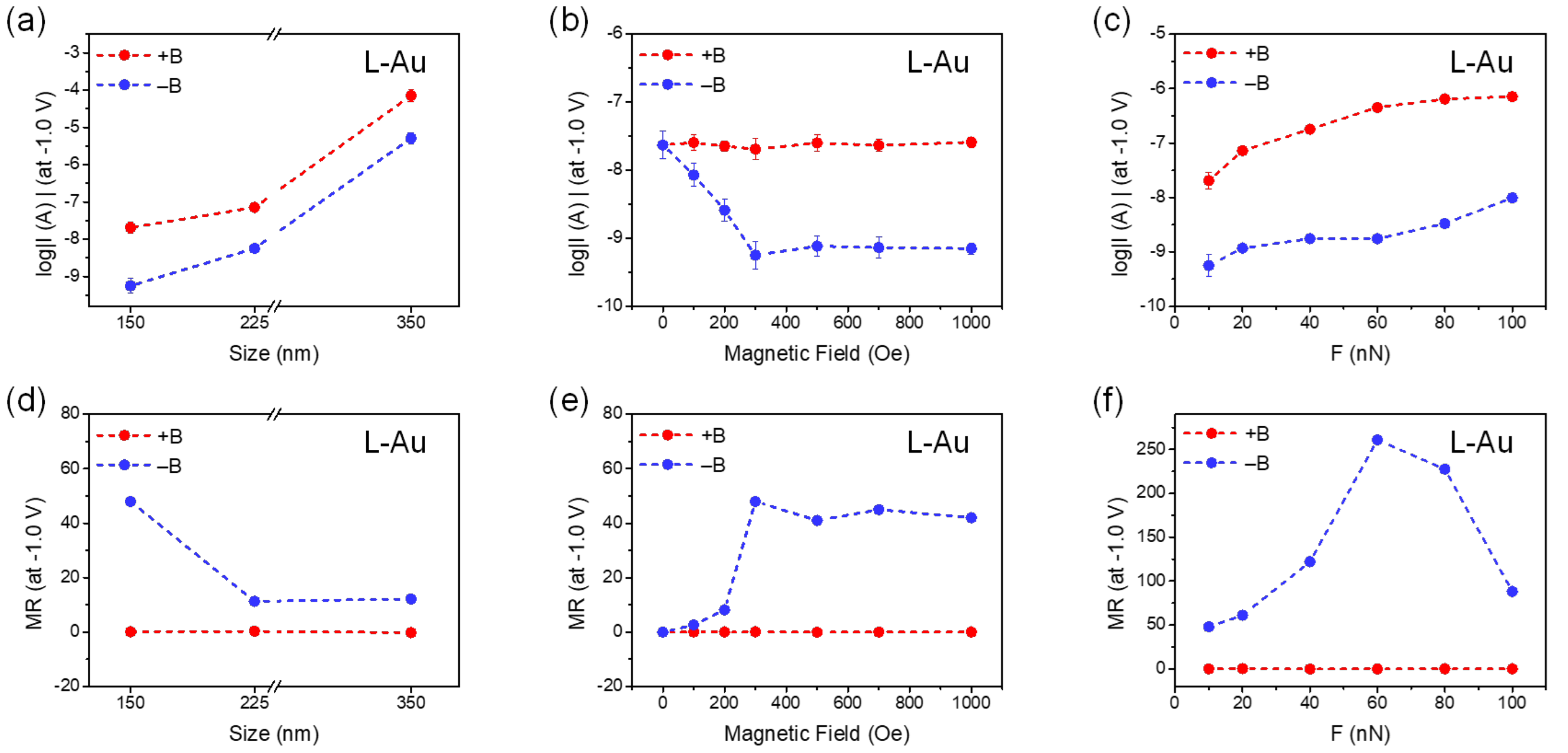}
\caption{Magnetoelectrical characterization of the ITO/L-Au NC/Pt junctions \textcolor{black}{measured by cAFM}. (a) and (d) Size-dependent currents and MR measured for L-Au NC. (b) and (e) Magnetic-field strength-dependent currents and MR of the ITO/L-Au NC/Pt junction. \textcolor{black}{(e) is also in Figure 2e as insert for better interpretation.} (d) and (f) The magnetoelectrical characterization of ITO/L-Au NC/Pt junctions as a function of the applied AFM tip force at the magnetic field direction of \textbf{+B} (red dashed line) and \textbf{–B} (blue dashed line). 
}
\label{exFigure9}
\end{figure}

\begin{figure}[tp!]
\centering
\includegraphics[width=1\linewidth]{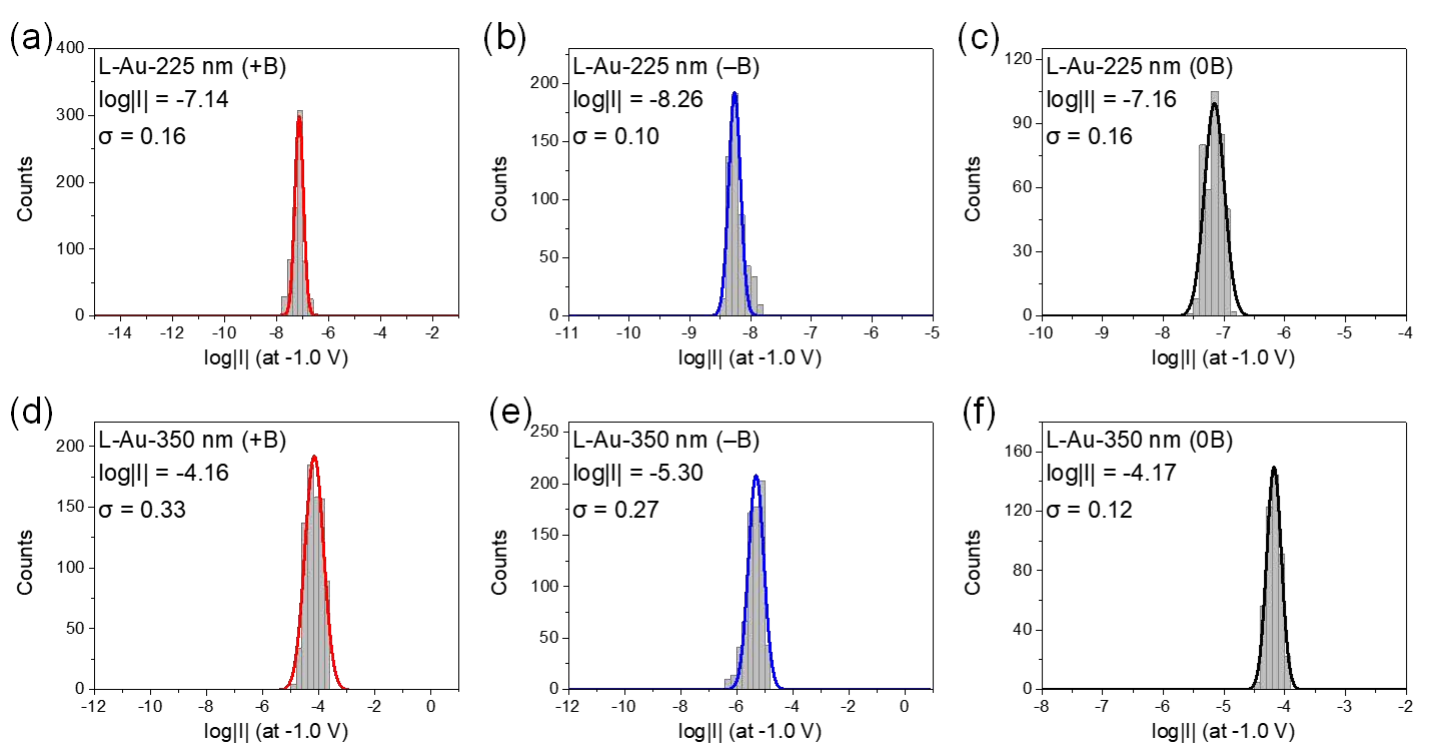}
\caption{Histograms of $log|I|$ for the L-Au NCs with different size: 225 nm (a-c) and 350 nm (d-f) under different directions of magnetic fields. These solid lines represent the Gaussian fits to these histograms. N = 390-690. \textcolor{black}{The I-V characterizations are performed by cAFM.}}
\label{exFigure10}
\end{figure}

\begin{figure}[tp!]
\centering
\includegraphics[width=1\linewidth]{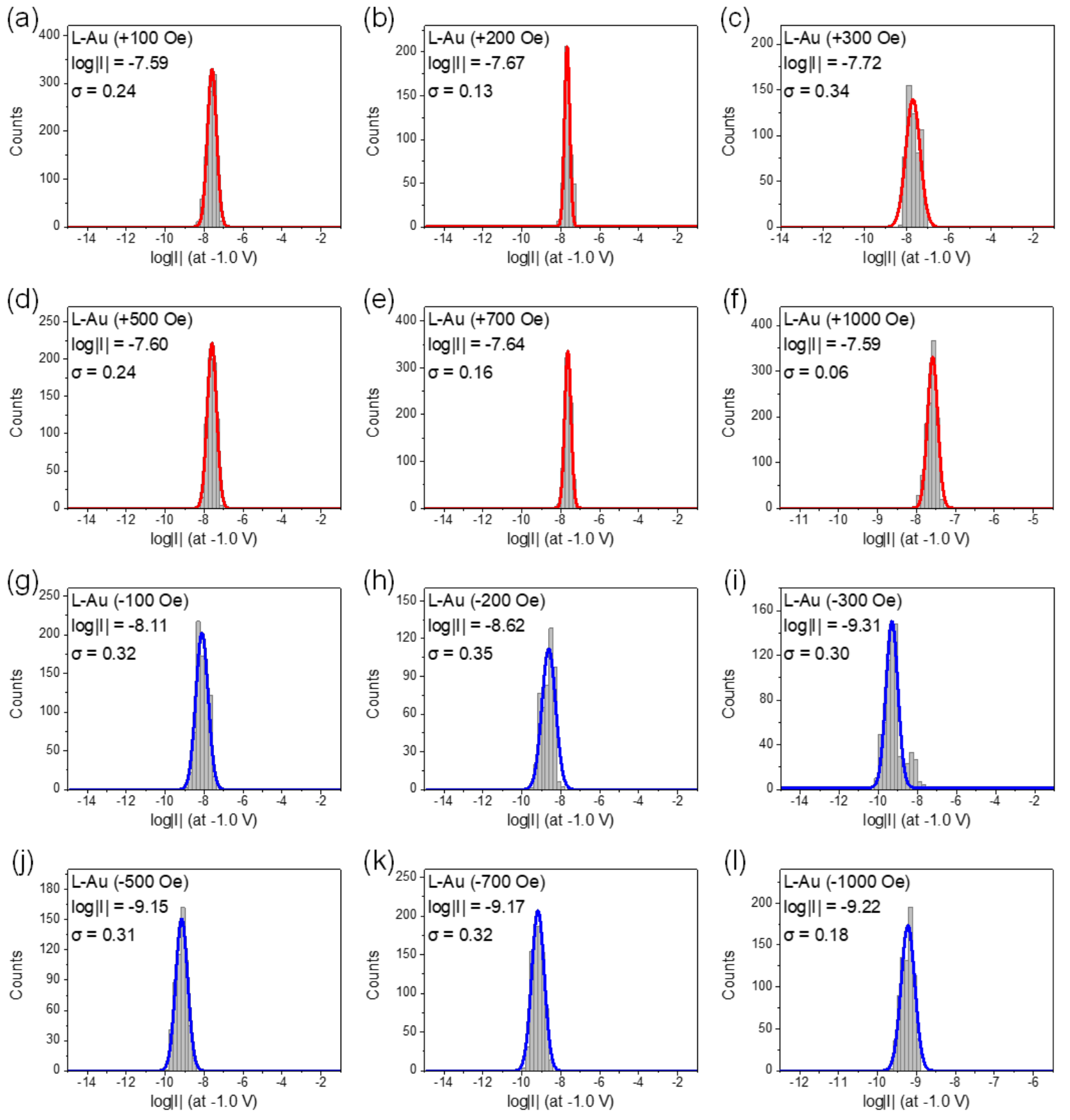}
\caption{
\textcolor{black} {Histograms of $log|I|$ with a Gaussian fit for the L-Au NC with the external magnetic field applied from +1000 Oe to –1000 Oe. (a, g) $\pm 100\,\text{Oe}$, (b, h) $\pm 200\,\text{Oe}$, (c, i) $\pm 300\,\text{Oe}$, (d, j) $\pm 500\,\text{Oe}$, (e, k) $\pm 700\,\text{Oe}$, (f, l) $\pm 1000\,\text{Oe}$. N = 478-973. The I-V characterizations are performed by cAFM.}}
\label{exFigure11}
\end{figure}

\begin{figure}[tp!]
\centering
\includegraphics[width=1\linewidth]{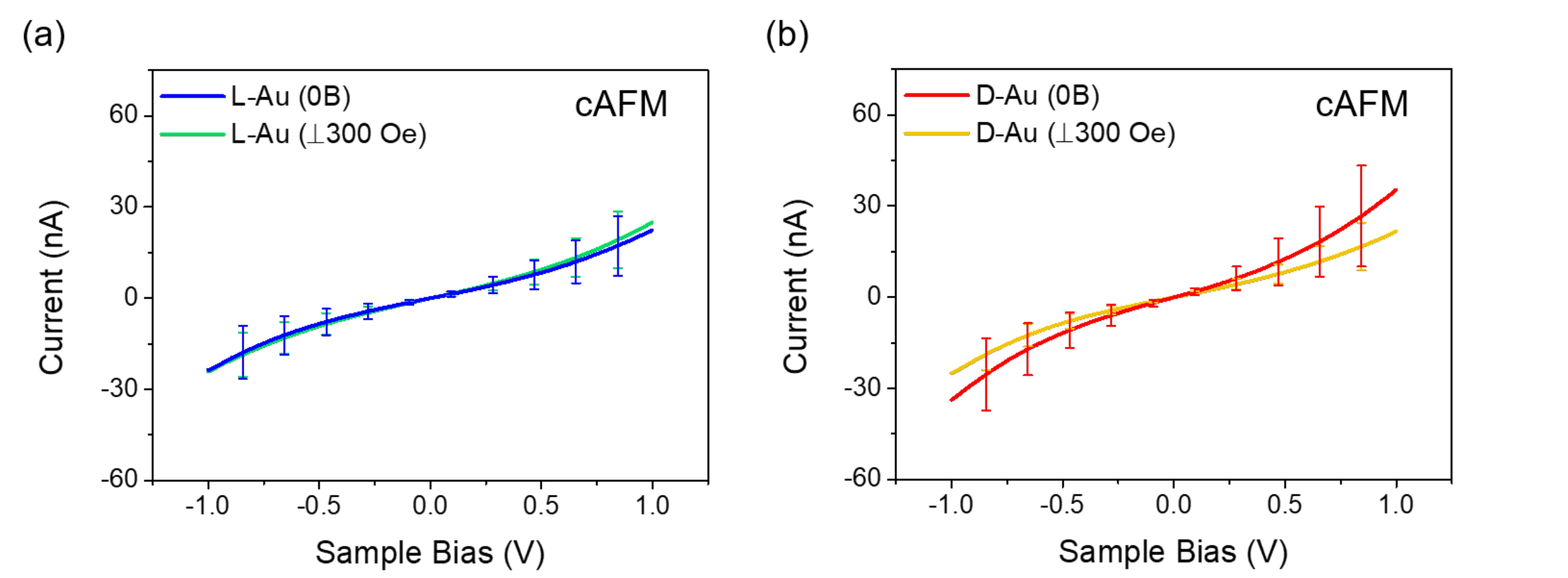}
\caption{\textcolor{black}{The current of ITO/chiral-Au NC/Pt junctions under the in-plane magnetic field. (a) and (b) I-V curves of the L-Au and D-Au NCs under magnetic field of \textbf{0B} and 300 Oe, respectively. The 300 Oe magnetic field is applied perpendicular to the current direction. The I-V characterizations are performed by cAFM.}}
\label{exFigure22}
\end{figure}

\begin{figure}[tp!]
\centering
\includegraphics[width=1\linewidth]{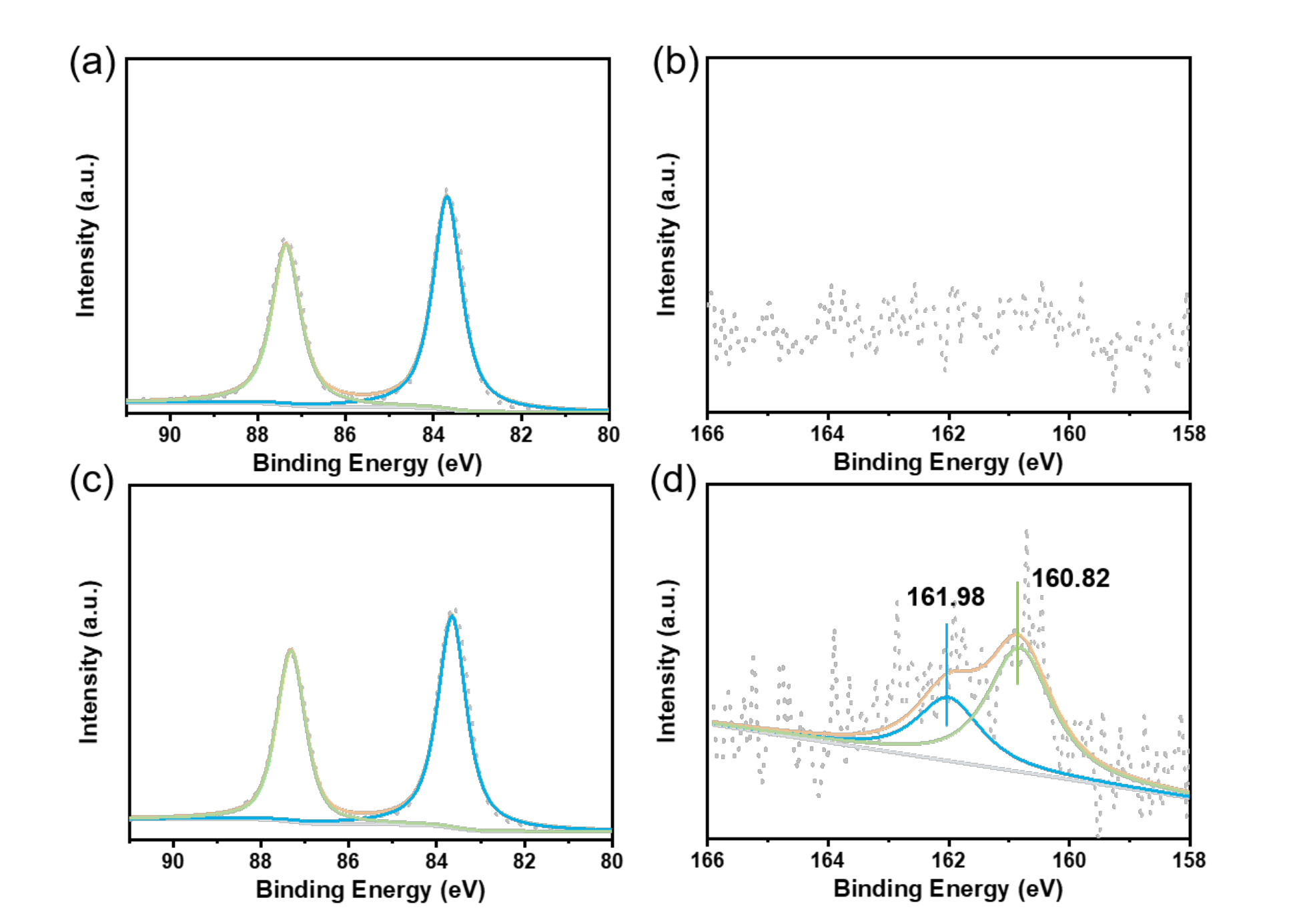}
\caption{
XPS spectra of chiral Au NCs and sulfide-modified chiral Au NCs (a) Au 4f region for D-Au NCs, (b) S 2p region for D-Au NCs, (c) Au 4f region for Sm-D-Au NCs, (d) S 2p region for Sm-D-Au NCs. D-Au NCs and Sm-D-Au NCs are taken as examples.}
\label{exFigure14}
\end{figure}

\begin{figure}[tp!]
\centering
\includegraphics[width=1\linewidth]{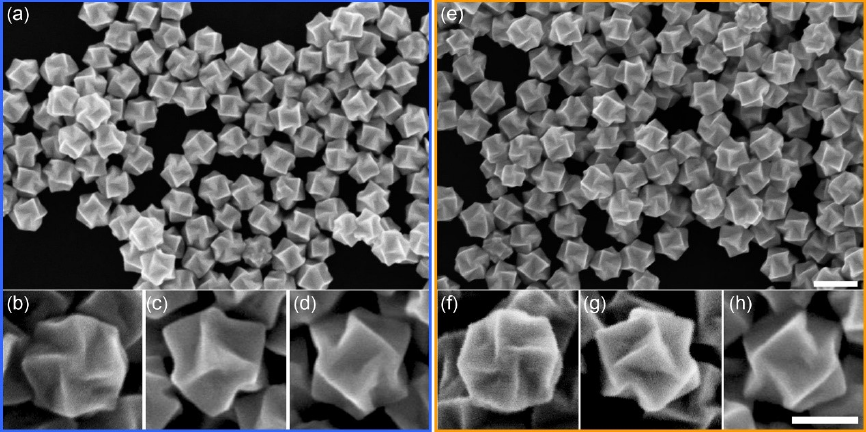}
\caption{SEM images of (a-d) sulfide-modified Au NCs: Sm-L-Au NCs (151.1±3.5 nm), (e-h) Sm-D-Au NCs (151.7±3.4 nm). Scale bars: 200 nm in (a) and (e); 100 nm in (b-d) and (f-h).}
\label{exFigure12}
\end{figure}

\begin{figure}[tp!]
\centering
\includegraphics[width=1\linewidth]{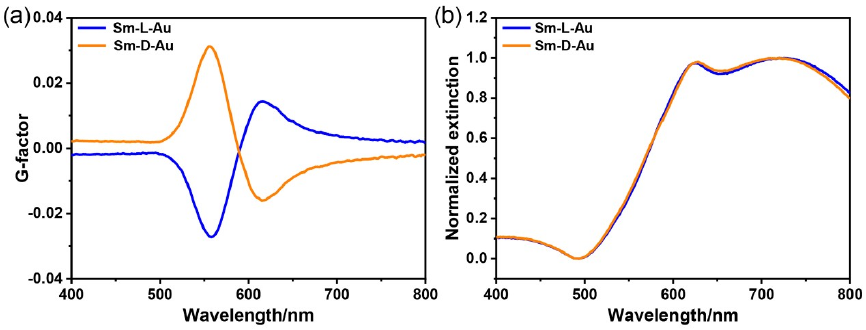}
\caption{(a) G-factor spectra and (b) Normalized extinction spectra of sulfide-modified chiral Au NCs.}
\label{exFigure13}
\end{figure}

\begin{figure}[tp!]
\centering
\includegraphics[width=1\linewidth]{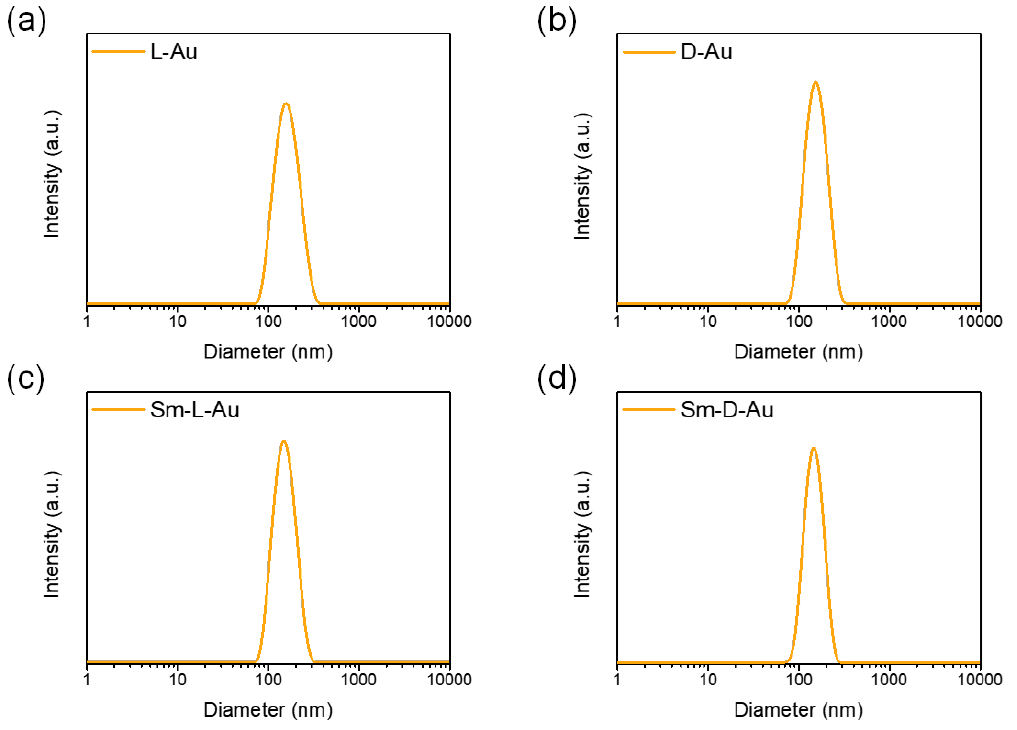}
\caption{Size (diameter/nm) distribution curves of chiral Au nanocrystals and sulfide-chiral Au nanocrystals (a) Size spectra of chiral Au nanocrystals from L-Cys, (b) Size spectra of chiral Au nanocrystals from D-Cys, (c) Size spectra of sulfide-chiral Au nanocrystals from L-Cys, (d) Size spectra of sulfide-chiral Au nanocrystals from D-Cys.}
\label{exFigure15}
\end{figure}

\begin{table}
    \centering
    \caption{Size and zeta potential distribution of chiral Au nanocrystals and sulfide-chiral Au nanocrystals by dynamic light scattering and zeta potential measurements.}
    \label{exTable3}
    
    \begin{tabular}{ccccc}
    \hline
Sample	& L-Au	& Sm-L-Au &	D-Au	& Sm-D-Au \\
\hline
Size(diameter)/nm &	152.94 &	146.06 &	149.46 &	144.12 \\
Zeta Potential/mV &	0.64 &	1.04 &	1.22 &	1.62 \\
        \hline
    \end{tabular}
\end{table}

 \begin{figure}[tp!]
\centering
\includegraphics[width=1\linewidth]{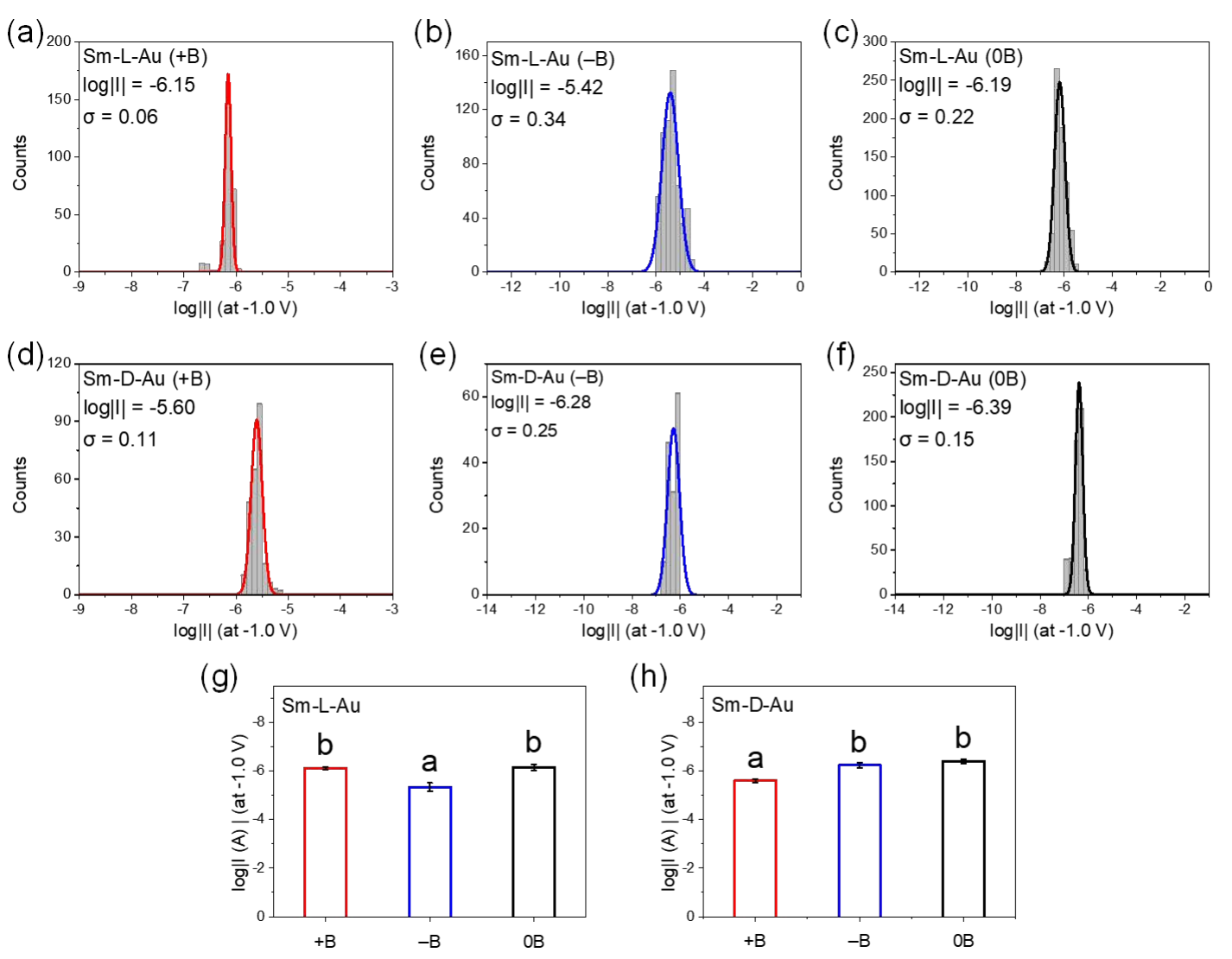}
\caption{Histograms of $log|I|$ with a Gaussian fit for the sulfide-modified chiral Au NCs: Sm-L-Au NCs (a-c) and Sm-D-Au NCs (d-f) under different directions of magnetic fields. N = 180-650. (g) and (h) Corresponding $log|I|$ at -1.0 V with statistical significance for Sm-L-Au and Sm-D-Au NCs under magnetic field directions of +\text{B} (red), –\text{B} (blue), and 0\text{B} (black). Different letters represent significant differences at the P $<$ 0.05 level based on one-way AVOVA. \textcolor{black}{The I-V characterizations are performed by cAFM.}
}
\label{exFigure17}
\end{figure}

\begin{figure}[tp!]
\centering
\includegraphics[width=1\linewidth]{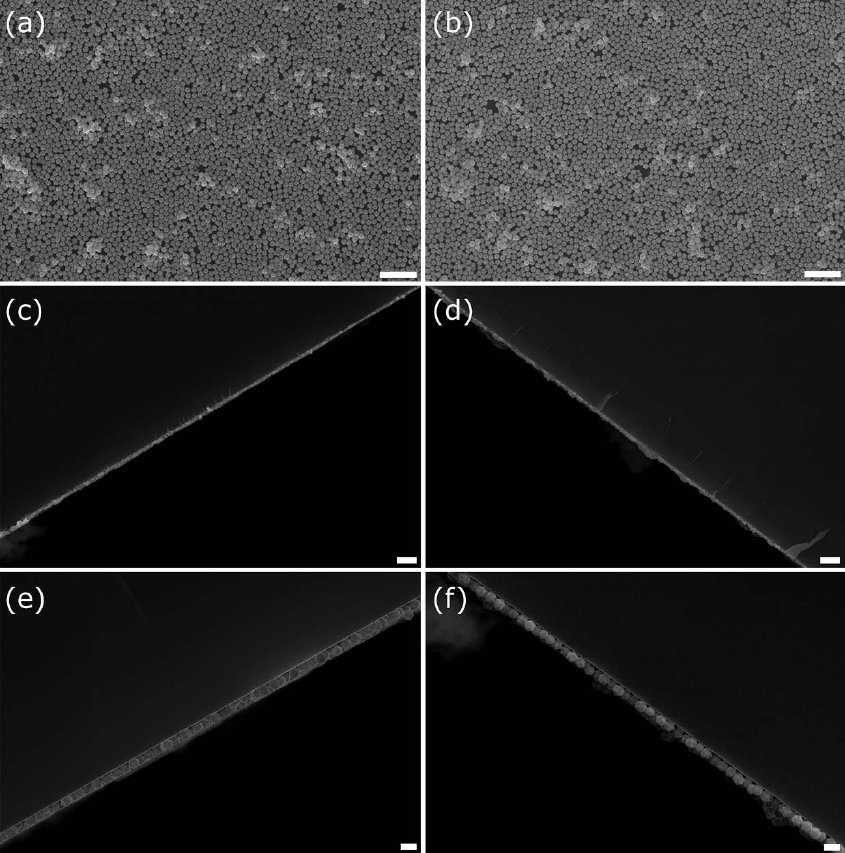}
\caption{\textcolor{black}{(a and b) Top view of SEM images of chiral Au NCs monolayers on Si substrates. (a) L-Au NCs monolayer. (b) D-Au NCs monolayer. (c-f) Side view of SEM images of chiral Au NCs monolayers on Si substrates. (c and e) L-Au NCs monolayer. (d and f) D-Au NC monolayer. The chiral Au NCs monolayers on stripped Au electrodes for solid-state device experiments are similar to those on Si substrates. Scale bars: 1~$\mu$m in a-d, 200 nm in e and f.}}
\label{exFigure28}
\end{figure}

\begin{figure}[tp!]
\centering
\includegraphics[width=0.8\linewidth]{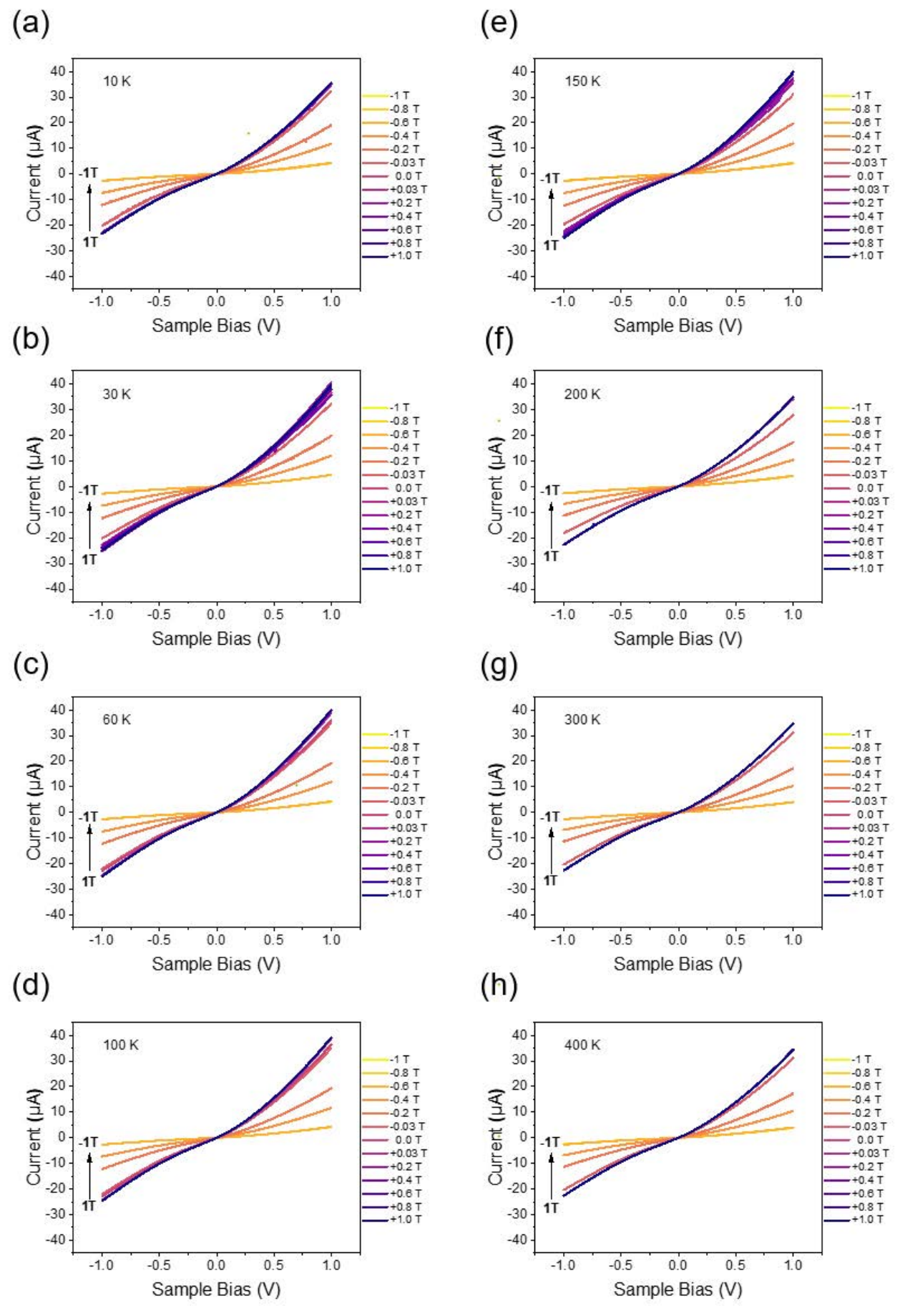}
\caption{\textcolor{black}{I-V curves of L-Au monolayer solid-state device at various temperature at (a) 10 K, (b) 30 K, (c) 60 K, (d) 100 K, (e) 150 K, (f) 200 K, (g) 300 K, and (h) 400 K. The magnetic field is from 1.0 T to -1.0 T.}}
\label{exFigure30}
\end{figure}

\begin{figure}[tp!]
\centering
\includegraphics[width=0.8\linewidth]{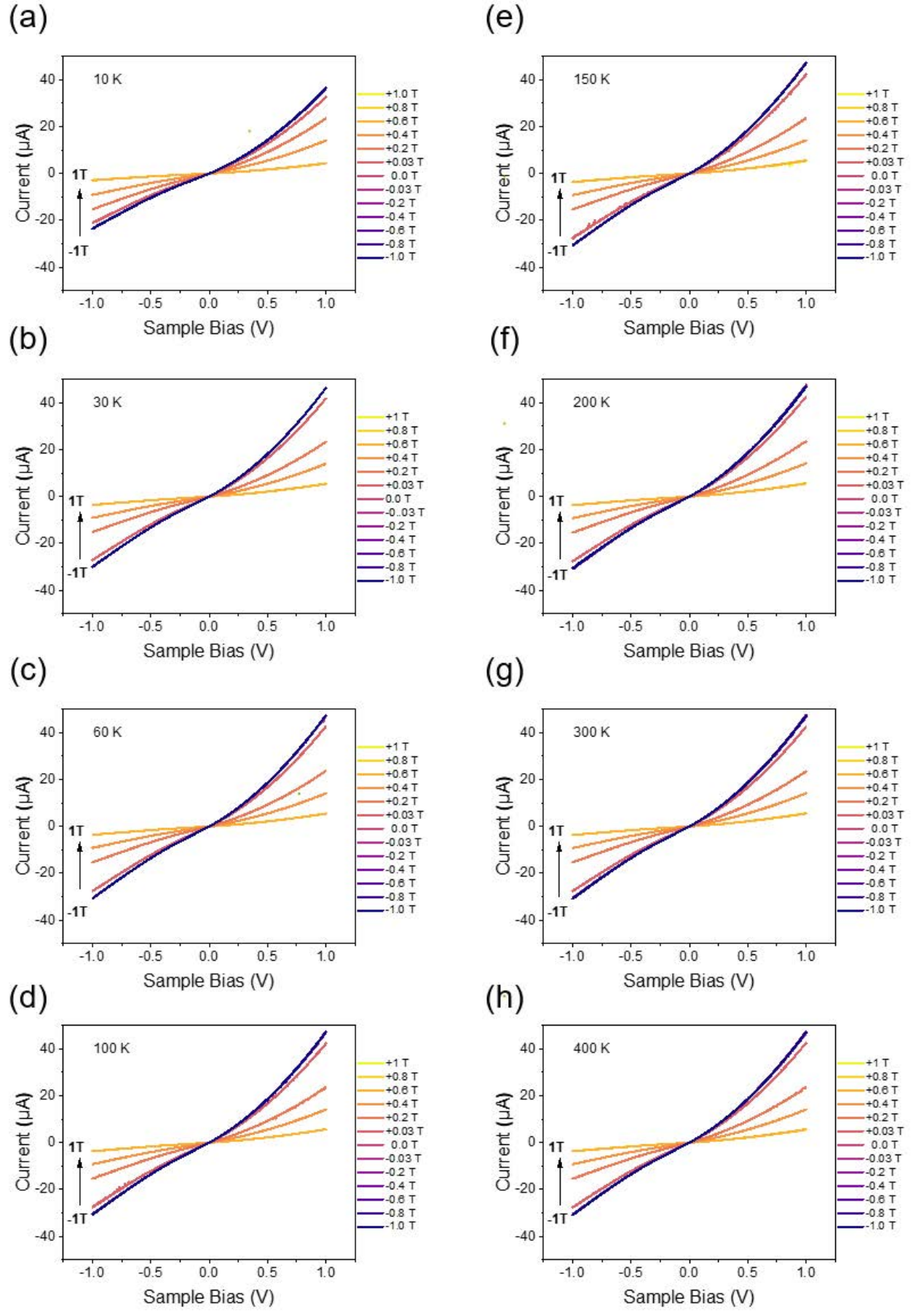}
\caption{\textcolor{black}{I-V curves of D-Au monolayer solid-state device at various temperature at (a) 10 K, (b) 30 K, (c) 60 K, (d) 100 K, (e) 150 K, (f) 200 K, (g) 300 K, and (h) 400 K. The magnetic field is from 1.0 T to -1.0 T.}}
\label{exFigure31}
\end{figure}

\begin{figure}[tp!]
\centering
\includegraphics[width=1\linewidth]{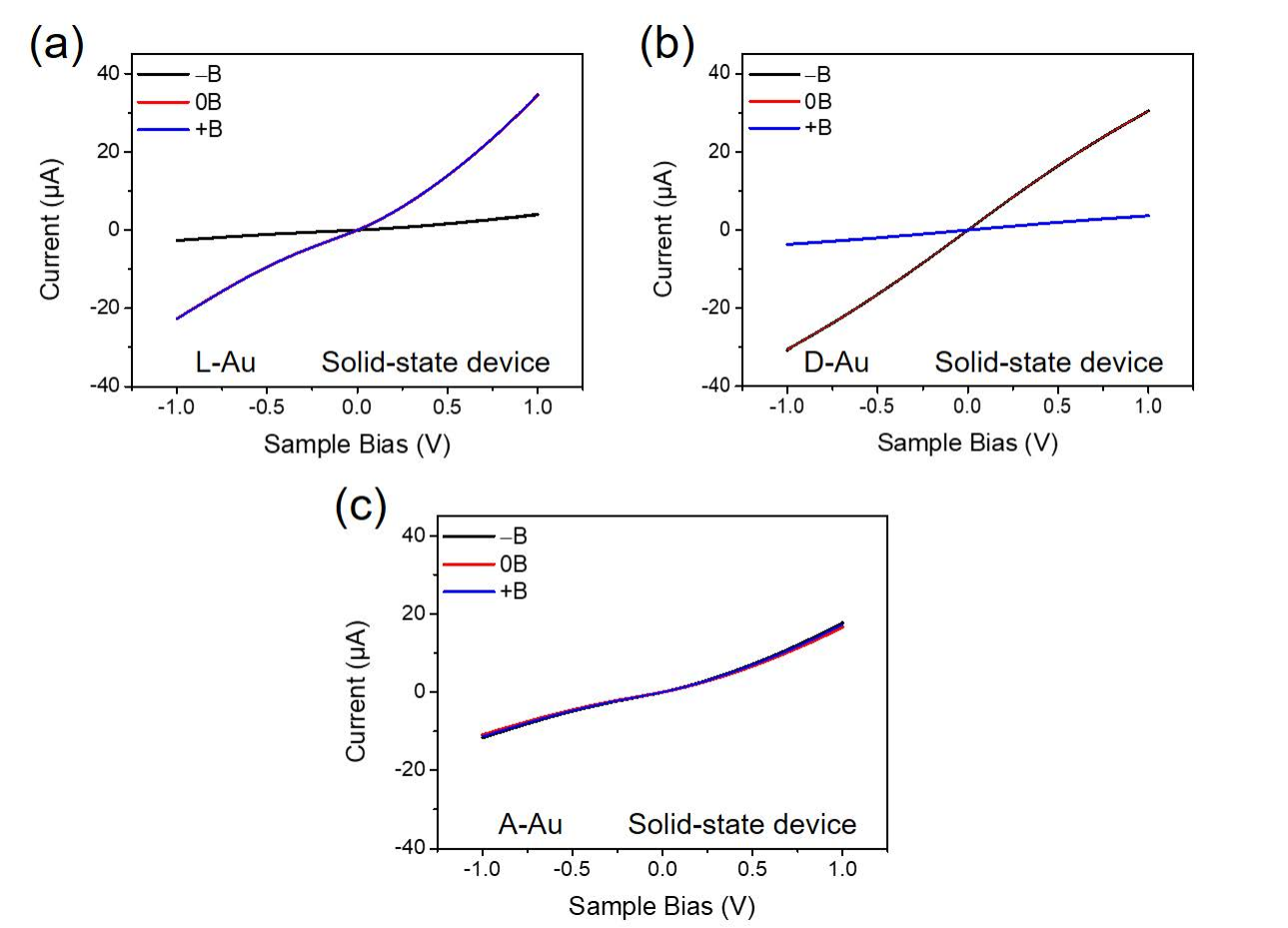}
\caption{\textcolor{black}{I-V curves of the Au/chiral Au NC monolayer/Au solid-state device under \textbf{+B}, \textbf{-B}, and \textbf{0B} at 300 K. (a) L-Au NC monolayer, (b) D-Au NC monolayer, (c) A-Au NC monolayer.}}
\label{exFigure24}
\end{figure}

\begin{figure}[tp!]
\centering
\includegraphics[width=1\linewidth]{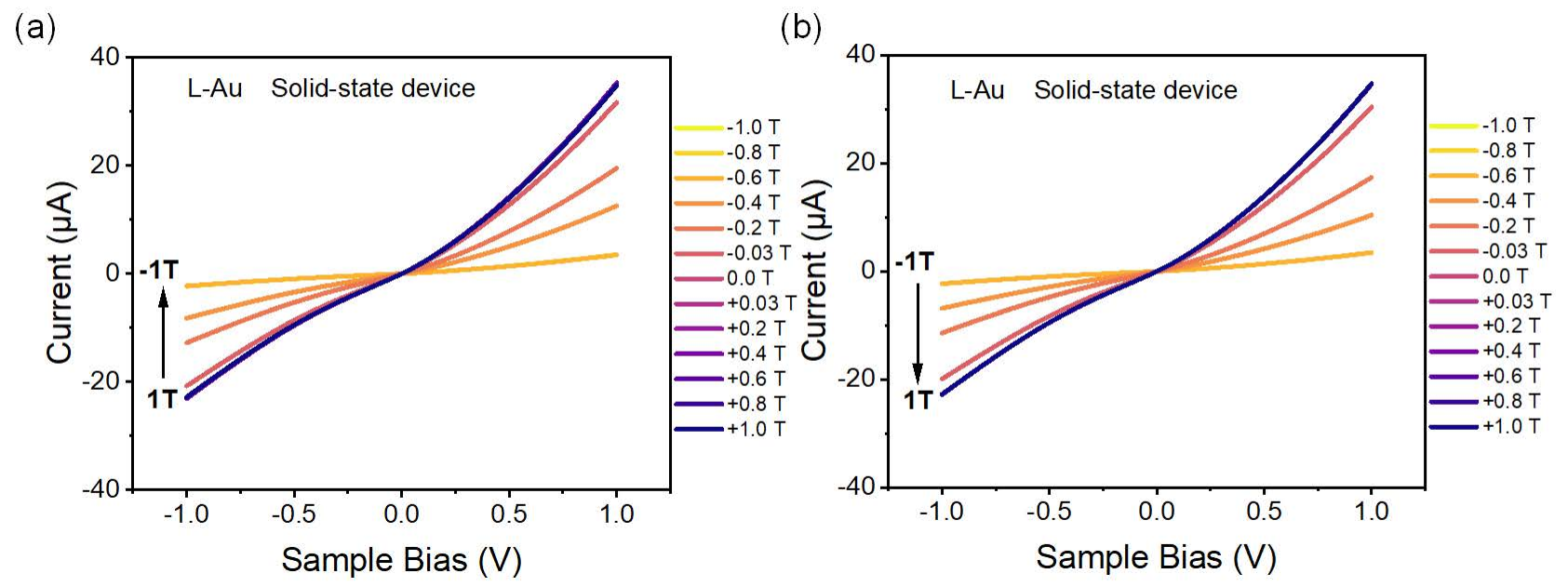}
\caption{\textcolor{black}{The I-V curves of L-Au monolayer in solid-state device measured under different orders of applied magnetic fields at 300 K: (a) applied magnetic fields from +1.0 T to -1.0 T, and (b) applied magnetic fields from -1.0 T to +1.0 T.}}
\label{exFigure26}
\end{figure}

\begin{figure}[tp!]
\centering
\includegraphics[width=0.5\linewidth]{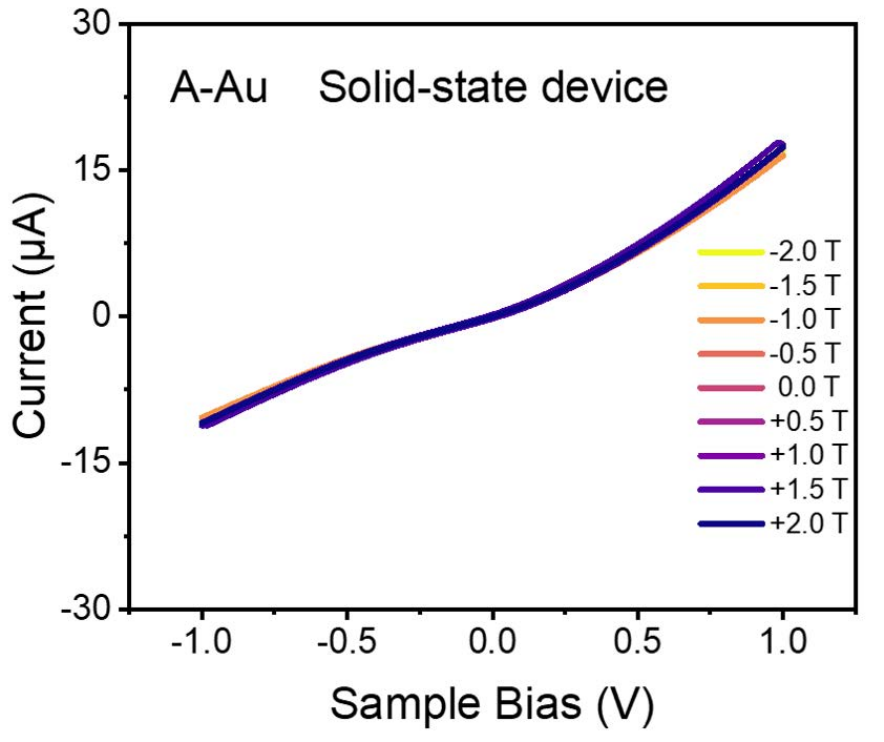}
\caption{\textcolor{black}{The I-V curves of achiral Au NC monolayer in solid-state device under magnetic fields from -2.0 to +2.0 T. It indicates that the novel MR is driven by chirality rather than possible magnetism in Au NCs.}}
\label{exFigure27}
\end{figure}

\begin{figure}[tp!]
\centering
\includegraphics[width=1\linewidth]{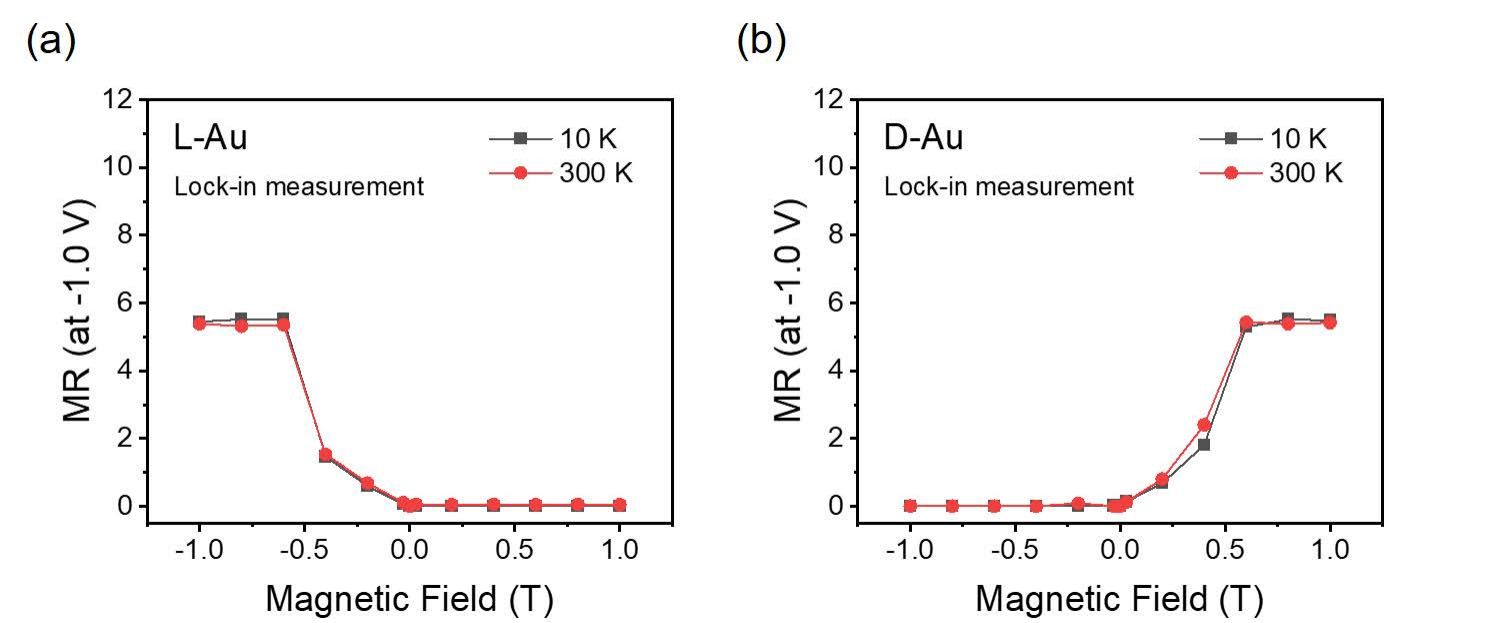}
\caption{\textcolor{black}{MR of the chiral Au NC monolayers in the solid-state devices measured by the 1st harmonic lock-in technique under magnet field ranging between -1.0 and 1.0 T. (a) L-Au, (b) D-Au. A 3 mV AC signal with a frequency of 177.1 Hz is applied on a -1.0 V DC bias for measurements. The MR values are measured at 10 K (black) and 300 K (red), respectively. }}
\label{exFigure29}
\end{figure}

\begin{figure}[tp!]
\centering
\includegraphics[width=1\linewidth]{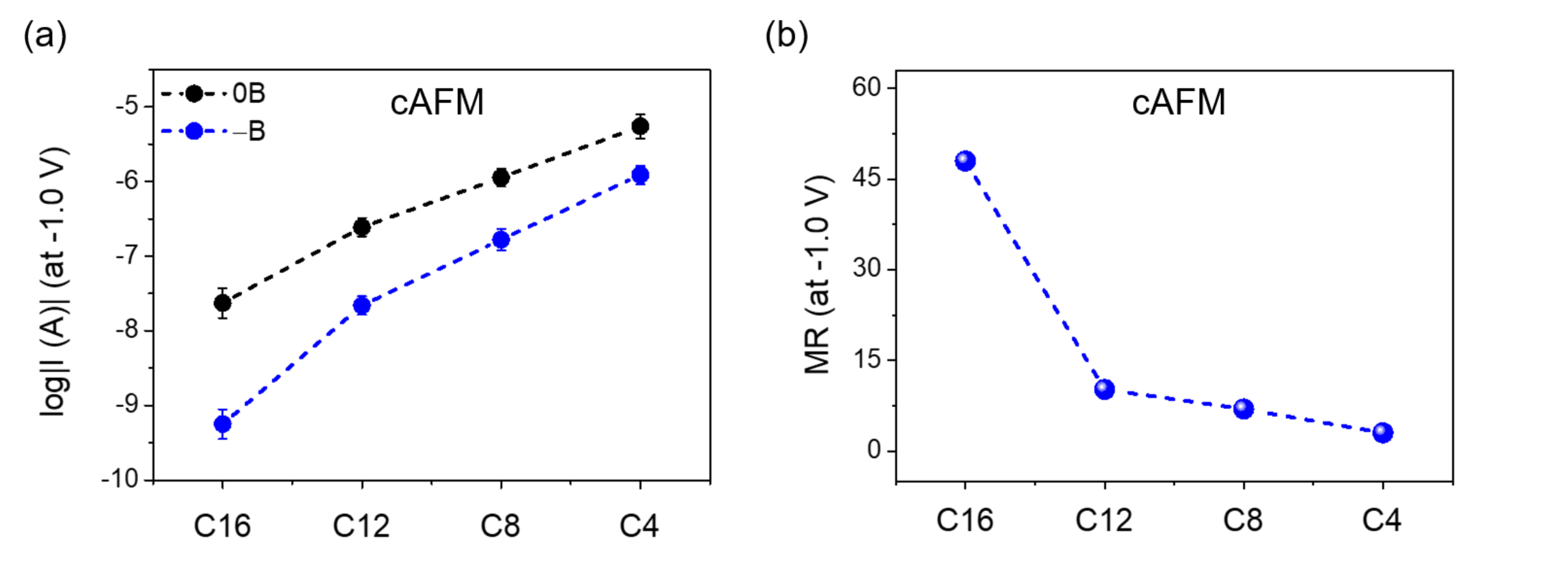}
\caption{\textcolor{black}{Magnetoresistance characterization of ITO/L-Au NC/Pt junctions with C$_n$TAB molecular shells of varying chain lengths (n=16, 12, 8, 4). (a) and (b) Shell molecular thickness-dependent currents and MR measured for the L-Au NC at the magnetic field directions of \textbf{-B} (blue) and \textbf{0B} (black). C$_4$TAB, C$_8$TAB, C$_{12}$TAB and C$_{16}$TAB correspond to CH$_3$(CH$_2$)$_3$N(CH$_3$)$_3$Br, CH$_3$(CH$_2$)$_7$N(CH$_3$)$_3$Br, CH$_3$(CH$_2$)$_{11}$N(CH$_3$)$_3$Br, and CH$_3$(CH$_2$)$_{15}$N(CH$_3$)$_3$Br, respectively. The I-V characterizations are performed by cAFM.}}
\label{exFigure23}
\end{figure}

 \begin{figure}[tp!]
\centering
\includegraphics[width=1\linewidth]{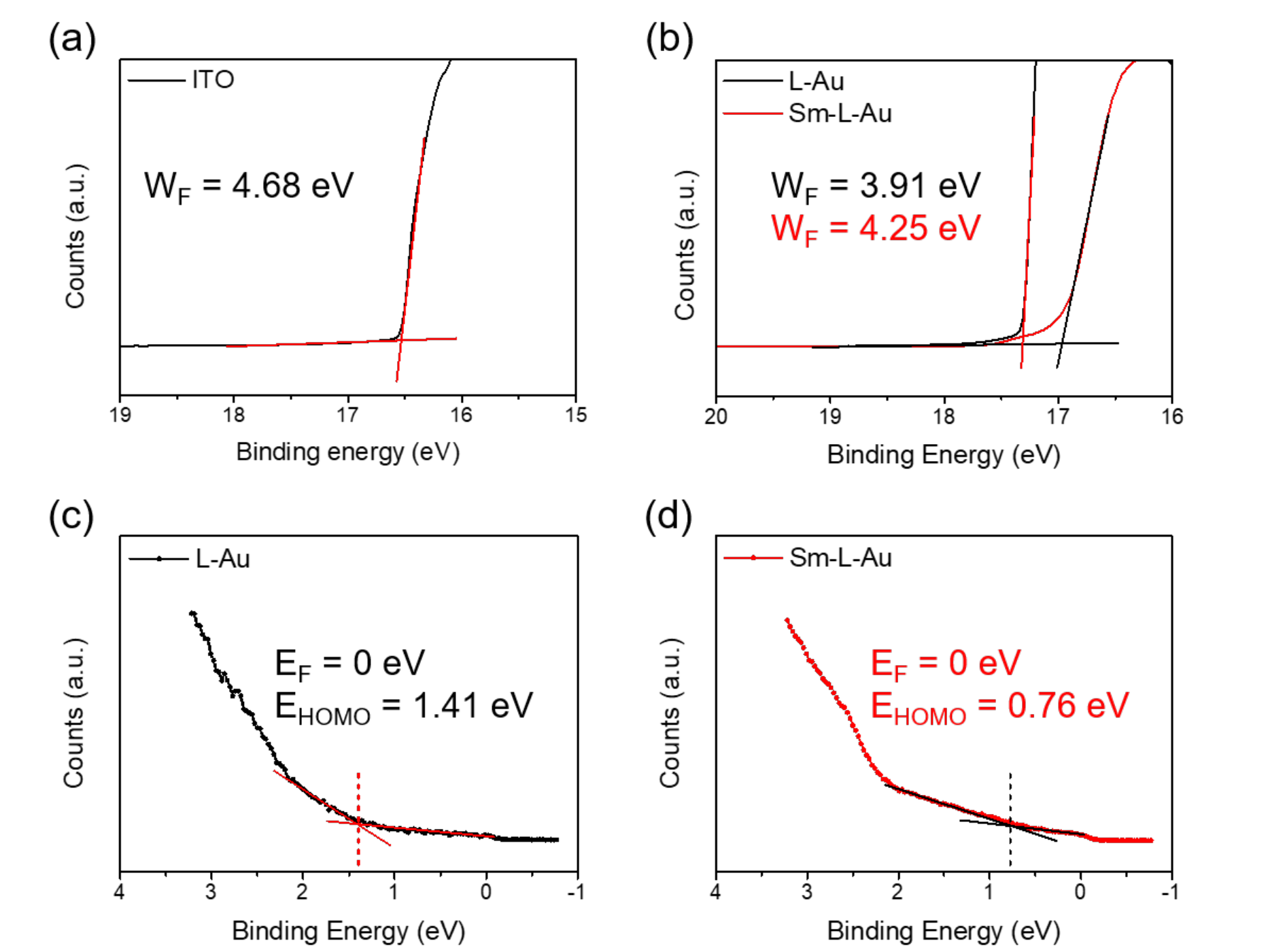}
\caption{UPS spectra of \textcolor{black}{ITO, }L-Au NCs, and Sm-L-Au NCs. (a)-(b) The work function ($W_F$) was determined by the secondary electron cut off. (b)-(c) The \textcolor{black}{highest occupied molecular orbital (HOMO)} of L-Au and Sm-L-Au, referring to the Fermi level ($E_{F} = 0$).}
\label{exFigure18}
\end{figure}

\begin{figure}[tp!]
\centering
\includegraphics[width=1\linewidth]{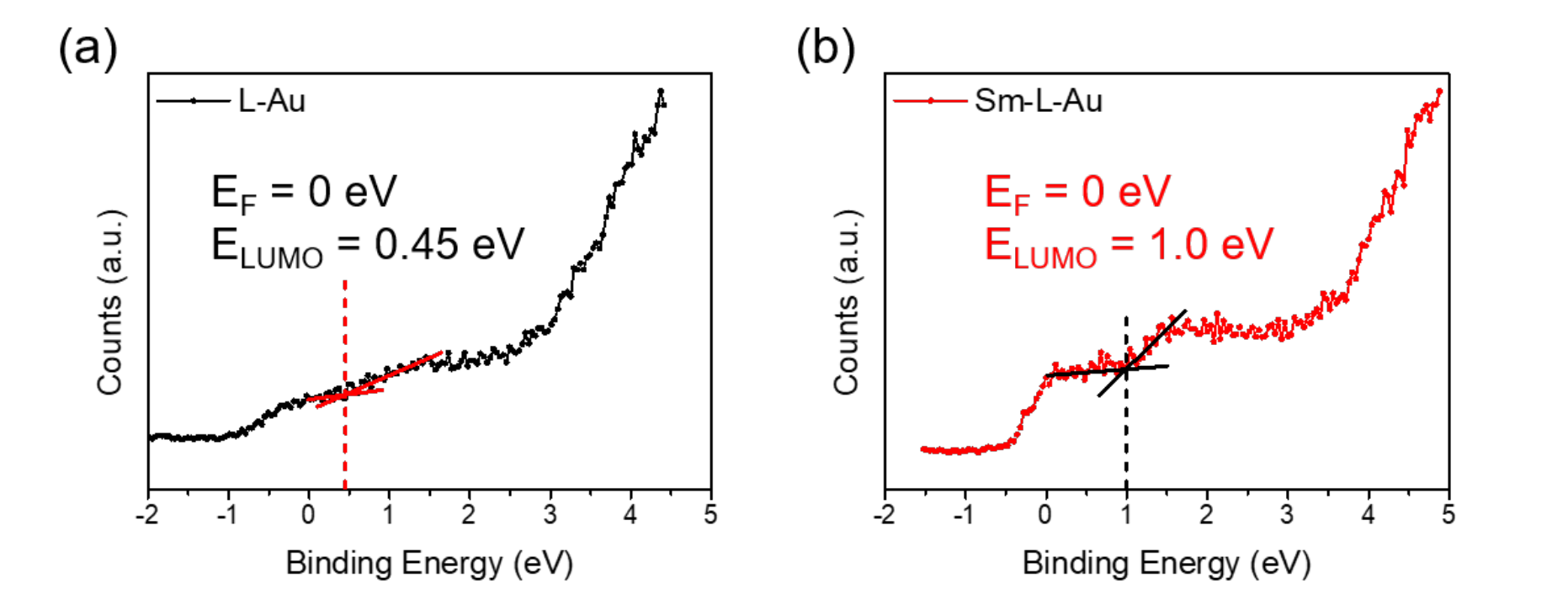}
\caption{The \textcolor{black}{lowest unoccupied molecular orbital (LUMO) }of L-Au NCs (a) and Sm-L-Au NCs (b) were measured by low energy inverse photoemission spectroscopy (LEIPS), when the \textcolor{black}{Fermi} level ($E_{F}$) is 0 \,\text{eV}.}
\label{exFigure19}
\end{figure}

\begin{figure}[tp!]
\centering
\includegraphics[width=1\linewidth]{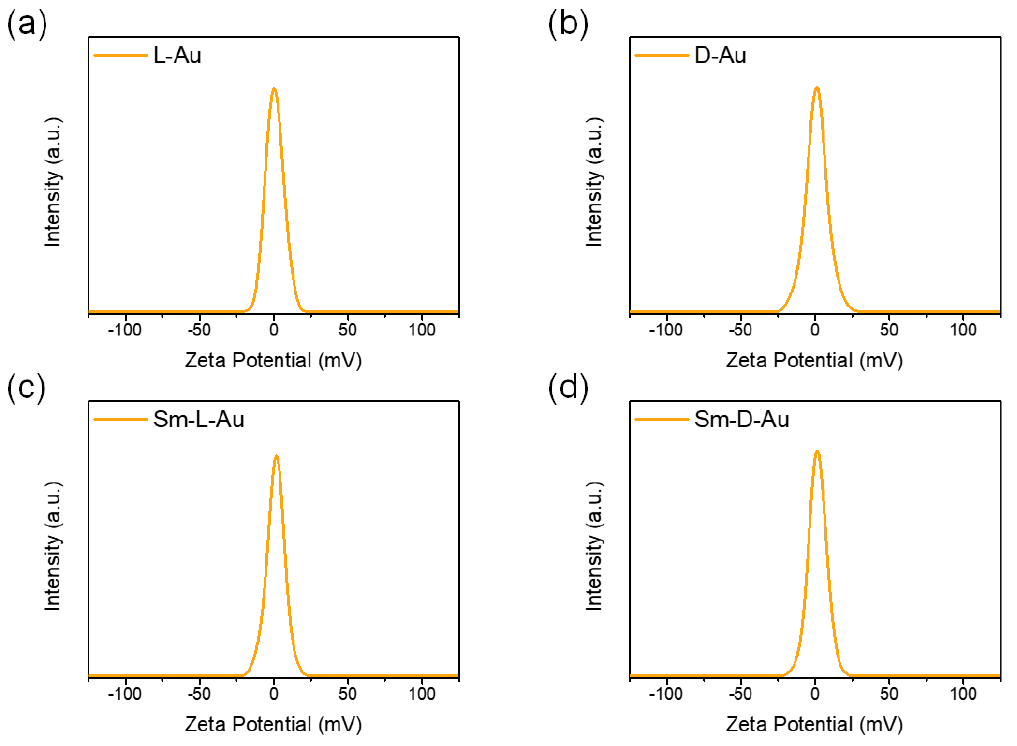}
\caption{Zeta potential (mV) distribution curves of chiral Au nanocrystals and sulfide-chiral Au nanocrystals (a) Zeta potential spectra of chiral Au nanocrystals from L-Cys, (b) Zeta potential spectra of chiral Au nanocrystals from D-Cys, (c) Zeta potential spectra of sulfide-chiral Au nanocrystals from L-Cys, (d) Zeta potential spectra of sulfide-chiral Au nanocrystals from D-Cys.}
\label{exFigure16}
\end{figure}

\begin{table}
\centering
\caption{Summary of the fitted parameters of Equation \ref{equation1} for the I-V curves in Figures~\ref{figure2},~\ref{figure3} and ~\ref{figure4}.}
\label{exTable4}

\hspace*{-1.5cm}
\begin{tabular}{cccccccccc}
\hline
                                                                          &    & \multicolumn{2}{c}{$U$} & \multicolumn{2}{c}{$A$}                               & \multicolumn{2}{c}{$\gamma$} & \multicolumn{2}{c}{$g$}                                 \\ 
                                                                          &    & Value    & St. Error  & Value                    & St. Error                & Value        & St. Error     & Value                     & St. Error                 \\ \hline
\multirow{3}{*}{\begin{tabular}[c]{@{}c@{}}L-Au \\ (Fig. 2)\end{tabular}} & \textbf{0B}  & 0.57349  & 0.02582     & \multirow{3}{*}{1.78039} & \multirow{3}{*}{0.03058} & \multirow{3}{*}{0.03395}     & \multirow{3}{*}{0.00226}       & \multirow{3}{*}{1.870E-7} & \multirow{3}{*}{2.420E-8} \\
                                                                          & \textbf{+B} & 0.57958   & 0.02575    &                          &                          &        &         &                           &                           \\
                                                                          & \textbf{-B} & 1.25512  & 0.03306    &                          &                          &        &         &                           &                           \\ 
                                                                          \\
\multirow{3}{*}{\begin{tabular}[c]{@{}c@{}}D-Au \\ (Fig. 2)\end{tabular}} & \textbf{0B}  & 0.60705  & 0.01150    & \multirow{3}{*}{1.85664} & \multirow{3}{*}{0.01004} & \multirow{3}{*}{0.01782}        & \multirow{3}{*}{0.00119}       & \multirow{3}{*}{3.513E-7} & \multirow{3}{*}{1.967E-8} \\
                                                                          & \textbf{+B} & 1.15073  & 0.00840    &                          &                          &        &         &                           &                           \\
                                                                          & \textbf{-B} & 0.58541  & 0.01162    &                          &                          &        &        &                           &                           \\  
                                                                          \\
\multirow{3}{*}{\begin{tabular}[c]{@{}c@{}}A-Au\\ (Fig. 3)\end{tabular}} & \textbf{0B} & 0.90373  & 0.02369    & \multirow{3}{*}{1.76905} & \multirow{3}{*}{0.00570} & \multirow{3}{*}{0.11616}     & \multirow{3}{*}{0.00064}       & \multirow{3}{*}{1.053E-7} & \multirow{3}{*}{7.316E-9}  \\
                                                                            & \textbf{+B} & 0.93746   & 0.02306     &                          &                          &       &         &                           &                            \\
                                                                            & \textbf{-B} & 0.83160   & 0.02468     &                          &                          &       &         &                           &                            \\
                                                                            \\
\multirow{3}{*}{\begin{tabular}[c]{@{}c@{}}Lm-A-Au\\ (Fig. 3)\end{tabular}} & \textbf{0B} & 1.42287  & 0.13744    & \multirow{3}{*}{1.49675} & \multirow{3}{*}{0.03990} & \multirow{3}{*}{-0.04132}     & \multirow{3}{*}{0.00106}       & \multirow{3}{*}{6.205E-8} & \multirow{3}{*}{1.362E-8}  \\
                                                                            & \textbf{+B} &  1.38262  &  0.13983    &                          &                          &       &         &                           &                            \\
                                                                            & \textbf{-B} &  1.52017  &  0.13019    &                          &                          &       &         &                           &                            \\
                                                                            \\
\multirow{3}{*}{\begin{tabular}[c]{@{}c@{}}Dm-A-Au\\ (Fig. 3)\end{tabular}} & \textbf{0B} & 0.70580  & 0.01353      & \multirow{3}{*}{1.78124}  & \multirow{3}{*}{0.00530} & \multirow{3}{*}{-0.13136}     & \multirow{3}{*}{0.00075}       & \multirow{3}{*}{1.337E-8} & \multirow{3}{*}{6.964E-10} \\
                                                                            & \textbf{+B} & 0.83243  & 0.01186     &                          &                          &       &         &                           &                            \\
                                                                            & \textbf{-B} & 0.66001  & 0.01383     &                          &                          &       &         &                           &                            \\
                                                                            \\
\multirow{3}{*}{\begin{tabular}[c]{@{}c@{}}Sm-L-Au\\ (Fig. 4)\end{tabular}} & \textbf{0B} & 0.98507  & 0.00402    & \multirow{3}{*}{1.85126} & \multirow{3}{*}{0.00852} & \multirow{3}{*}{-0.11099}    & \multirow{3}{*}{0.00129}       & \multirow{3}{*}{3.110E-5} & \multirow{3}{*}{9.513E-7}  \\
                                                                            & \textbf{+B} & 0.97131  & 0.00383    &                          &                          &       &         &                           &                            \\
                                                                            & \textbf{-B} & 0.53743  & 0.00553    &                          &                          &       &        &                           &                            \\
                                                                            \\
\multirow{3}{*}{\begin{tabular}[c]{@{}c@{}}Sm-D-Au\\ (Fig. 4)\end{tabular}} & \textbf{0B} & 0.97719  & 0.00460    & \multirow{3}{*}{1.85991} & \multirow{3}{*}{0.00837} & \multirow{3}{*}{-0.14875}      & \multirow{3}{*}{0.00153}       & \multirow{3}{*}{1.880E-5} & \multirow{3}{*}{7.072E-7}  \\
                                                                            & \textbf{+B} & 0.56409  & 0.00772    &                          &                          &      &       &                           &                            \\
                                                                            & \textbf{-B} & 0.92600  & 0.00467    &                          &                          &       &         &                           &                           \\
                                                                          \hline
\end{tabular}
\end{table}

\end{document}